\documentclass[a4paper,11pt,english]{article}
\pdfoutput=1

\usepackage{ifdraft}
\ifdraft{\newcommand{\authnote}[3]{{\color{#3} {\bf  #1:} #2}}}{\newcommand{\authnote}[3]{}}
\newcommand{\anote}[1]{\authnote{ András}{#1}{green}}

\usepackage[utf8]{inputenc}
\usepackage[T1]{fontenc}
\usepackage[left=1in,top=1in,right=1in,bottom=1in]{geometry} 
\usepackage[english]{babel}
\usepackage{verbatim}

\usepackage[dvipsnames]{xcolor} 
\usepackage{tikz}
\usepackage{tikz-cd}
\usepackage{amssymb}
\usepackage{mathtools}
\usepackage{bbm}
\usepackage{mleftright}\mleftright
\usepackage{multirow}
\usepackage{amsthm} 
\theoremstyle{plain}
\usepackage{thmtools} 
\usepackage{thm-restate}

\usepackage{float}
\usepackage{enumitem}

\usepackage{caption}
\usepackage{subcaption}

\usepackage{titling}
\usepackage[final=true,colorlinks = true,allcolors = {blue},]{hyperref}
\usepackage[capitalise]{cleveref}

\newcommand{\eps}{\varepsilon}

\newcommand{\ketbra}[2]{|#1\rangle\! \langle #2|}
\newcommand{\braketbra}[3]{\langle #1|#2| #3 \rangle}
\newcommand{\nrm}[1]{\left\lVert #1 \right\rVert}
\newcommand{\bigO}[1]{\mathcal{O}\left( #1 \right)}
\newcommand{\bigOt}[1]{\widetilde{\mathcal{O}}\left( #1 \right)}

\DeclarePairedDelimiter\bra{\langle}{\rvert}
\DeclarePairedDelimiter\ket{\lvert}{\rangle}
\DeclarePairedDelimiterX\braket[2]{\langle}{\rangle}{#1 \delimsize\vert #2}
\newcommand{\underflow}[2]{\underset{\kern-60mm \overbrace{#1} \kern-60mm}{#2}}

\def\Pr{\mathrm{Pr}}

\providecommand{\poly}[1]{\mathrm{poly}\left(#1\right)}

\newtheorem{theorem}{Theorem}
\newtheorem{corollary}[theorem]{Corollary}
\newtheorem{lemma}[theorem]{Lemma}

\newtheorem{definition}[theorem]{Definition}

\newtheorem*{claim*}{Claim}

\usepackage{tikz,pgfplots,tkz-graph}\pgfplotsset{compat=1.16}
\usetikzlibrary{backgrounds,fit,decorations.pathreplacing,calc,arrows}

\title{
	(Sub)Exponential advantage of \\ adiabatic quantum computation with no sign problem
}
\author{
	András Gilyén\thanks{Institute for Quantum Information and Matter, California Institute of Technology. \tt{agilyen@caltech.edu}}
	\and
	Umesh Vazirani\thanks{Berkeley Quantum Information \& Computation Center, UC Berkeley. \tt{vazirani@cs.berkeley.edu}}
}
\date{\today\vspace{-5mm}}

\begin{document}
	
	\maketitle
	
	\begin{abstract}
		We demonstrate the possibility of (sub)exponential quantum speedup via a quantum algorithm that follows an adiabatic path of a gapped Hamiltonian with no sign problem. This strengthens the superpolynomial separation recently proved by Hastings~\cite{hastings2020PowerOfAdiabaticNoSign}. The Hamiltonian that exhibits this speed-up comes from the adjacency matrix of an undirected graph, and we can view the adiabatic evolution as an efficient $\mathcal{O}(\mathrm{poly}(n))$-time quantum algorithm for finding a specific "EXIT" vertex in the graph given the "ENTRANCE" vertex. On the other hand we show that if the graph is given via an adjacency-list oracle, there is no classical algorithm that finds the "EXIT" with probability greater than $\exp(-n^\delta)$ using at most $\exp(n^\delta)$ queries for $\delta= \frac15 - o(1)$. 
		Our construction of the graph is somewhat similar to the ``welded-trees'' construction of Childs et al.~\cite{childs2003ExpSpeedupQW}, but uses additional ideas of Hastings~\cite{hastings2020PowerOfAdiabaticNoSign} for achieving a spectral gap and a short adiabatic path. 
	\end{abstract}
	
	\section{Introduction}

	Adiabatic quantum computing~\cite{farhi2000QCompAdiabatic} is an interesting model of computation that is formulated directly in terms of Hamiltonians, the quantum analog of constraint satisfaction problems (CSPs). The computation starts in the known ground state of an initial Hamiltonian, and slowly (adiabatically) transforms the acting Hamiltonian into a final Hamiltonian whose ground state encapsulates the answer to the computational problem in question. The final state of the computation is guaranteed, by the quantum adiabatic theorem, to have high overlap with the desired ground state, as long as the running time of the adiabatic evolution is polynomially large in the inverse of the smallest spectral gap of any Hamiltonian along the adiabatic path~\cite{ambainis2004ElementaryProofQAdiabThm}. This model has been intensely studied, not only because of its inherent interest, but also because it is the zero-temperature limit of quantum annealing. 

	In full generality, adiabatic quantum computing is known to be equivalent to standard circuit-based quantum computing~\cite{aharonov2007AdiabQCompEquivStandQComp}. A very interesting question is what is the power of adiabatic quantum computing where all Hamiltonians were "stoquastic," i.e., restricted to not having a sign problem. What this means is that in some basis all off-diagonal terms of $H$ are non-positive. Adiabatic quantum computing with no sign problem includes the most natural case where the final Hamiltonian is diagonal, and represents the objective function to be optimized, and the initial Hamiltonian consists of Pauli $X$ operators acting on each qubit, with ground state the uniform superposition on all the $n$-bit strings. This question was also motivated by understanding the computational limits of the quantum annealers implemented by the company D-Wave, where all the Hamiltonians were stoquastic. 
	
	Bravyi and Terhal~\cite{bravyi2010ComplexityOfStoqFrustFreeHam} showed that for frustration-free Hamiltonians without a sign problem, computing the ground state is classically tractable,
	thereby raising the question of whether this was true for general Hamiltonians without a sign problem. Indeed, a stronger conjecture was that quantum Monte-Carlo, a widely used heuristic in computational condensed matter physics, already provided a technique for an efficient classical simulation. This latter possibility was ruled out by a result of Hastings and Freedman~\cite{hastings2013ObstructionsToClSimQAdiabaticAlg}, who showed the existence of topological obstructions to the convergence of quantum Monte Carlo on such problems. 

	The question of classical tractability for general Hamiltonians with no sign problem was open until it was addressed in a recent breakthrough by Hastings~\cite{hastings2020PowerOfAdiabaticNoSign}, who showed a superpolynomial oracle separation between classical algorithms and adiabatic quantum computation with no sign problem. In this paper, we heavily build on Hastings' ideas but extend his results in two ways. First, we show that there is a (sub)exponential oracle separation, of the form $2^{n^{\delta}}$ between classical algorithms and adiabatic quantum computation with no sign problem. Our second contribution is conceptual. Our construction of the Hamiltonian is fairly transparent -- it consists of a graph with an ENTRANCE and an EXIT vertex, and the challenge is: given the ENTRANCE vertex and oracle access to the adjacency matrix of the graph, find the EXIT vertex. A simple quantum walk finds the EXIT vertex in polynomial time, and likewise so does a simple adiabatic algorithm which carries out a straight line interpolation between the initial and final Hamiltonian. 
	
	Our simple construction highlights the similarities and differences with the well known ``welded-trees" graph (see \cref{fig:evSolutionsPlot} and \cref{sec:weldedWalk}) which is the basis of the first known example of exponential speedup by quantum walks~\cite{childs2003ExpSpeedupQW}. The welded-trees graph is not suitable for adiabatic computation, since the ground state has exponentially small support on the roots of the two trees (the ENTRANCE and EXIT vertices). To see this, notice that a quantum walk on the welded trees may be viewed as walking on the symmetric subspace of each level of the trees -- i.e. it is just a path of length $2 \mathrm{depth} + 1$. This path has uniform edge weights, except at the middle edge, which has $\sqrt{2}$-times bigger weight. This makes the largest eigenvector of the path graph decay exponentially from the middle towards the two ends. The starting point of our construction is the simple observation that equalizing the edge weights in the level graph on symmetric subspaces has the effect of fixing the exponential decay problem. On the other hand, this necessarily makes the underlying graph non-regular. In turn, this enables classical algorithms to detect the structure of the graph~\cite{childs2020GraphPropExpQSpeedup,benDavid2020SymmetriesGraphPropertiesQSpeedups} and ultimately destroys the lower bound of~\cite{childs2003ExpSpeedupQW} that heavily build on the regularity of the graph. In order to restore the classical hardness result we use the approach of Hastings, who ensures hardness by ``decorating'' the graph by means of attaching a cleverly shaped forest to every vertex. Another feature that can be seen very concretely in our simplified construction is the role of the $\ell_2$ versus $\ell_1$ normalization difference in the behavior of the quantum vs. classical walk. 
	
 	\section{Main results}		
		
		We follow the general framework from Hastings' paper. The main idea there was to start with a graph that the adiabatic algorithm can traverse efficiently, and to hide that graph within a larger graph as follows: attach a number of trees to each vertex of the original graph, so that the attached trees form the bulk of the new graph. Now, the intuition is that the behavior of a quantum walk versus classical walk on the attached trees would be governed by their $\ell_2 \to \ell_2$ (i.e., spectral) norm versus $\ell_1 \to \ell_1$ norm respectively, and the latter is quadratically larger. As a result the attached trees only negligibly affect the $\ell_2$-weight distribution of the ground state (and so quantum algorithms only suffer from a minor perturbation), while they dramatically shift the $\ell_1$-weight distribution of the ground state away from the original graph. Intuitively speaking this enables the trees to lure away classical random walks from the original graph, so that they get lost in the attached ``camouflage trees'' with very high probability. Furthermore, by choosing the trees to have a confusing enough shape, one can ensure that there is no classical algorithm that can avoid getting drawn into the ``camouflage trees.'' Therefore, classical algorithms fail to quickly explore the original graph, and in our case this ultimately leads to their inability of efficiently finding the EXIT vertex. 
	
    	The classical hardness is achieved by constructing hard-to-navigate trees with a fractal-like structure that are built in a recursive manner, via a sequence of so-called ``decorations''~\cite{hastings2020PowerOfAdiabaticNoSign}. Hastings' decorations rapidly blew up the degrees of the vertices in the graph, thereby limited the number of subsequent decoration rounds to logarithmic, permitting only a superpolynomial separation. We on the other hand use a modified decoration sequence allowing polynomially many rounds of decoration, and ultimately leading to a (sub)exponential separation.
		
	\subsection{The basic adiabatic path at the core of our quantum algorithm}
	
	We begin with a simple underlying problem of starting at one endpoint of a path on $\ell$ vertices, and finding the other endpoint of the path.\footnote{We work with undirected and unweighted graphs, but for simplicity allow parallel edges and self-loops. One can think about parallel edges as simple integer edge weights, since they are represented in the same way in the adjacency matrix. Ultimately we will only use self-loops at the two distinguished vertices "ENTRANCE" and "EXIT".} A simple adiabatic algorithm for this problem is specified as follows: Let $A_\ell$ denote the adjacency matrix of the path $\braketbra{k}{A_\ell}{k+1}=\braketbra{k+1}{A_\ell}{k}=1$, and let the corresponding Hamiltonian be $H_\ell:=-A_\ell$, while $H^{(i)}:=-\ketbra{1}{1}$ and $H^{(f)}:=-\ketbra{\ell}{\ell}$.
	
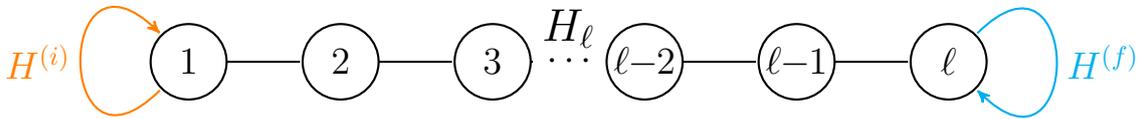
	\begin{center}
		\begin{tikzpicture}[->,>=stealth',shorten >=1pt,thick]
		\SetGraphUnit{2} 
		\tikzset{VertexStyle/.style = {draw,circle,thick,
				minimum size=1cm,
				font=\Large\bfseries},thick,EdgeStyle/.style = {-}} 
		\Vertex[L={$1$}]{1} \EA[L={$2$}](1){2} \EA[L={$3$}](2){3} \EA[L={$\kern-1.7mm\ell\kern-1.3mm-\kern-1.3mm2\kern-1.7mm$}](3){l-2} \EA[L={$\kern-1.7mm\ell\kern-1.3mm-\kern-1.3mm1\kern-1.7mm$}](l-2){l-1} \EA[L={$\ell$}](l-1){l}
		\Edge(1)(2)
		\Edge(2)(3)
		\Edge[label={\Large$\underset{\phantom{\scalebox{1.2}{$H_\ell$}}}{\overset{\scalebox{1.2}{$H_\ell$}}{\kern.1mm\cdots}}$}, style = {}](3)(l-2)
		\Edge(l-2)(l-1)
		\Edge(l-1)(l)		
		
		\Loop[dist=2cm,dir=WE,label=\Large$H^{(i)}$,labelstyle=left,color=orange](1)  
		\Loop[dist=2cm,dir=EA,label={\Large$H^{(f)}$},labelstyle=right,color=cyan](l)  
		\end{tikzpicture} 
	\end{center}

	Consider the simple adiabatic path $H_\ell(s)$ that first interpolates between $H_i$ and $H_\ell$, then between $H_\ell$ and $H_f$, so that $H_\ell(s):=(1+s)H_\ell-s H^{(i)}$ for $s\in[-1,0]$ and $H_\ell(s):=(1-s)H_\ell+s H^{(f)}$ for $s\in[0,1]$.

	
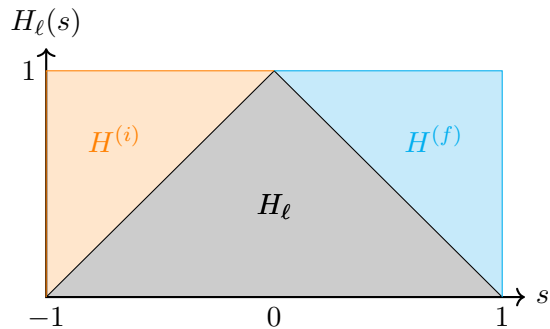
\begin{center}
		\begin{tikzpicture}[scale=3]
			\fill[fill=orange!20] (-1,0)--(-1,1)--(0,1);
			\node[color=orange] at (-0.7,0.7) {$H^{(i)}$};				
			\fill[fill=cyan!20] (0,1)--(1,1)--(1,0);
			\node[color=cyan] at (0.7,0.7) {$H^{(f)}$};				
			\fill[fill=black!20] (-1,0)--(0,1)--(1,0);
			\node[color=black] at (0,0.4) {$H_\ell$};				
			\draw [<->,thick] (-1,1.1) node (yaxis) [above] {$H_\ell(s)$}
			|- (1.1,0) node (xaxis) [right] {$s$};		
			\node[below] (s-1) at (-1,0) {$-1$};	
			\node[below] (s0) at (0,0) {$0$};	
			\node[below] (s1) at (1,0) {$1$};	
			\node[left] (h) at (-1,1) {$1$};	
			\draw[color=black] (-1,0)--(0,1)--(1,0);
			\draw[color=orange] (0,1)--(-1,1) -- (-1,0);
			\draw[color=cyan] (0,1)--(1,1) -- (1,0);
			\node[color=black] at (0,0.4) {$H_\ell$};		
							
		\end{tikzpicture} 
	\end{center}

	If one moves slowly enough along this adiabatic path~\cite{farhi2000QCompAdiabatic,ambainis2004ElementaryProofQAdiabThm}, the quantum evolution maps ``ENTRANCE''$:=\ket{1}$ -- the \emph{initial} ground state of $H^{(i)}$ to ``EXIT''$:=\ket{\ell}$ -- the \emph{final} ground state of $H^{(f)}$, since $H(s)$ has a gap of size $\Omega(\frac{1}{\ell^2})$ for all $s\in [-1,1]$, see \cref{sec:pathGap}. Note that if one wishes to use only a simple ``straight'' adiabatic path, and stops at $s=0$, a measurement in the computational basis still reveals the state $\ket{\ell}$ with probability at least $\Omega(\ell^{-3})$ since the ground state of $H_\ell$ has $\Omega(\ell^{-\frac32})$ overlap with $\ket{\ell}$, cf. \cref{sec:pathGap}.\footnote{Alternatively we could increase the success probability to $\Omega(1)$ by accepting any vertex in $\{\ell/2, \ell/2+1,\ldots, \ell\}$, thereby effectively defining multiple exits. This task can be made classically hard as well, similarly to the single EXIT scenario.} 
		
	\subsection{Making the task of finding EXIT classically hard}
	
	In order to prove classical hardness we will hide the EXIT vertex in a larger graph -- the new graph will be chosen to allow the quantum adiabatic algorithm to still be efficient, while making the task of any classical algorithm very difficult. The id's of the vertices will be chosen randomly in order to remove any non-structural hints about the whereabouts of the EXIT vertex, and the graph will be specified by oracle access to its adjacency list\footnote{Our graph has $N=\bigO{\exp(\poly{m})}$ vertices, each having at most $d=\max(5m,m^2+2m+1)$ neighbors. The classical adjacency-list oracle $O\colon [N]\times [d]\mapsto [N]\cup \{\star\}$ can be queried with the id of a vertex and a number~$k$, and as a response tells the $k$-th neighbor of the vertex with the given id (the neighbors are sorted arbitrarily). If the vertex has less than $k$ neighbors (with multiplicity), then the oracle outputs $\star$ as a response. We assume that the corresponding \emph{reversible} quantum oracle acts as $\ket{i}\ket{k}\ket{0}\to \ket{i}\ket{O(i,k)}\ket{n(i,k)}$, where $n(i,k)$ is the number of $h\in [k-1]$ such that $O(i,h)=O(i,k)$. This is a so-called \emph{in-place} adjacency-list oracle~\cite{berry2015HamSimNearlyOpt,gilyen2018QSingValTransf}, which can save us a $\poly{m}$ factor in the number of queries used by our quantum algorithms.}, together with the ENTRANCE vertex -- one of the two vertices with a self-loop.\footnote{Since there are only two vertices with a self-loop, and we know the ENTRANCE vertex we can simulate both the initial and the final Hamiltonians by using the adjacency-list oracle.} The task is to find the EXIT vertex -- the other vertex with a self-loop attached. The graph will have polynomially bounded maximal vertex degree, so the adiabatic evolution can be efficiently performed by a quantum computer using (time-dependent) sparse Hamiltonian simulation techniques~\cite{berry2019TimeDependentHamSimL1}.
	
	In order to make the task of finding EXIT classically hard we ``blow-up'' the path graph of length $\ell$ via two main modifications, that we call \emph{obfuscation} and \emph{decoration}.
	
	\begin{definition}[Obfuscation of a path of length $\ell$]\label{def:obf} 
		We replace every vertex that has distance\footnote{We use the notation $[n]:=\{1,2,3,\ldots,n\}$.} $d\in[k]$ from terminal vertices $\{$ENTRANCE, EXIT$\}$ by a cluster $C$ of $m^{2d}$ vertices and call these the \emph{funnel} vertices, and replace the other middle vertices (that have distance $d>k$) by a cluster of $m^{2k}$ vertices, and call those the \emph{tunnel} vertices. Then we add edges between clusters $C_j$ and $C_{j+1}$ corresponding to neighbor vertices $j$ and $j+1$ in $P_\ell$, so that we build an $m^2$-ary tree (with the terminal vertices as roots) on the funnel vertices. Between clusters that correspond to vertices with distance $d\geq k$ we add edges along $m$ random matchings. Additionally, in order to preserve spectral properties we add $2m$ self-loops to the ENTRANCE and the EXIT vertices, and an independently chosen random uniform degree-($4\cdot m$) expander graph on each cluster $C_j\colon j\in[\ell]\setminus\{1,\ell\}$, as in \cref{sec:expanders}.\footnote{We use random expander graphs as in \cref{def:randReg}, but condition on their spectral gap being at least $2m$. For large enough $m$ the effect of conditioning is negligible as shown by \cref{cor:ourExpanders}.} 
	\end{definition}
		
	Note that the graph on the tunnel vertices is $4m$-regular. The decoration construction, described next, will hang $m$ trees from each vertex of the obfuscated graph, each of them being a complete $(5m-1)$-ary tree (by a complete tree we mean a tree for which every node has the same number of children except at the bottom layer, which is at a fixed depth) on its first $\poly{m}$ layers, and then having gradually less children at later layers. The construction is motivated by its effect on the tunnel --- it will increase the degree of each tunnel vertices to $5m$. Thus, the resulting graph will still be $5m$-regular on the original tunnel vertices, as well as on the surrounding vertices in the first $\poly{m}$ layers of the added trees. This will make it very difficult for any classical algorithm to distinguishing edges between the tunnel vertices from edges that lead away from the tunnel, thereby making the traversal of the tunnel very slow. 
	
	If we would everywhere add a complete $(5m-1)$-ary tree of depth $d$, then the decoration trees would be easy to detect: after traversing an edge perform a non-backtracking walk of length $d$, if one arrives at a leaf it means that the traversed edge is hanging a decoration tree. Since we cannot add more than $\exp(\poly{m})$ new vertices, the trees must have a bounded depth. Therefore, in order to circumvent such detection algorithms we should construct trees where the distribution of the lengths before a non-backtracking random walk hits the bottom of a tree looks approximately self-similar, i.e., after going one level deeper in the tree the expected distribution should not change by more than a (sub)exponentially small amount. In order to achieve this, the decoration is carried out in $r=m^\delta$ rounds, giving the attached trees a complex fractal-like structure.
		
	\begin{definition}[Decoration]\label{def:dec}
		Let $G=(V,E)$ be a graph. A level-$j$ decoration graph $G_j$ is obtained from $G$ by ``decorating'' every vertex $v\in V$ by attaching $m^{(1-\delta)}$ new trees via an edge to their root. The attached trees are complete $(5m-(j-1)m^{(1-\delta)}-1)$-ary trees with depth $j m^{(3\delta + o(1))}$. We define $G^{(r)}$ as the $r$-round decoration of $G$, which is obtained from $G$ by applying a level-$r$ decoration, then subsequently level-$(r-1)$, level-$(r-2)$, $\ldots$, level-$1$ decorations.
	\end{definition}

	The above modified definition of decoration is the key to our improved separation result. Hastings~\cite{hastings2020PowerOfAdiabaticNoSign} added an increasing number of decoration trees in every round so that decoration doubled the maximal degree of the graph in each round. The rapid growth of the vertex degrees prohibited applying more than logarithmically many rounds of decoration, which ultimately limited the separation to being at most superpolynomial. In contrast, we only add $m^{1-\delta}$ decoration trees in each round, which enables us applying $m^\delta$ rounds of decoration, ultimately resulting in a (sub)exponential classical lower bound. In order to keep the desired increase in classical hardness despite using fewer decoration trees, we make them slightly deeper.  

\begin{figure}[ht]
	$\phantom{.}$\kern-8mm
	\begin{tikzpicture}[scale=1]
	\SetGraphUnit{1.5}
	\GraphInit[vstyle=Simple]
	\tikzset{VertexStyle/.style = {shape = circle,fill = black,minimum size = 3.5pt,inner sep=1pt}}
	
	\Vertex[x=-6,y=0]{N}
	
	\Vertex[x=-4.5,y= 3]{M1}
	\Vertex[x=-4.5,y= 1]{M2}
	\Vertex[x=-4.5,y=-1]{M3}
	\Vertex[x=-4.5,y=-3]{M4}
	\Vertex[x=-3,y=3.75]{M11}
	\Vertex[x=-3,y=3.25]{M12}
	\Vertex[x=-3,y=2.75]{M13}
	\Vertex[x=-3,y=2.25]{M14}
	\Vertex[x=-3,y=1.75]{M21}
	\Vertex[x=-3,y=1.25]{M22}
	\Vertex[x=-3,y=0.75]{M23}
	\Vertex[x=-3,y=0.25]{M24}
	\Vertex[x=-3,y=-0.25]{M31}
	\Vertex[x=-3,y=-0.75]{M32}
	\Vertex[x=-3,y=-1.25]{M33}
	\Vertex[x=-3,y=-1.75]{M34}
	\Vertex[x=-3,y=-2.25]{M41}
	\Vertex[x=-3,y=-2.75]{M42}
	\Vertex[x=-3,y=-3.25]{M43}
	\Vertex[x=-3,y=-3.75]{M44}		
	
	\Vertex[x=4.5,y= 3]{H1}
	\Vertex[x=4.5,y= 1]{H2}
	\Vertex[x=4.5,y=-1]{H3}
	\Vertex[x=4.5,y=-3]{H4}
	\Vertex[x=3,y=3.75]{H11}
	\Vertex[x=3,y=3.25]{H12}
	\Vertex[x=3,y=2.75]{H13}
	\Vertex[x=3,y=2.25]{H14}
	\Vertex[x=3,y=1.75]{H21}
	\Vertex[x=3,y=1.25]{H22}
	\Vertex[x=3,y=0.75]{H23}
	\Vertex[x=3,y=0.25]{H24}
	\Vertex[x=3,y=-0.25]{H31}
	\Vertex[x=3,y=-0.75]{H32}
	\Vertex[x=3,y=-1.25]{H33}
	\Vertex[x=3,y=-1.75]{H34}
	\Vertex[x=3,y=-2.25]{H41}
	\Vertex[x=3,y=-2.75]{H42}
	\Vertex[x=3,y=-3.25]{H43}
	\Vertex[x=3,y=-3.75]{H44}
	
	\Vertex[x=-1,y=3.75]{T1}
	\Vertex[x=-1,y=3.25]{T2}
	\Vertex[x=-1,y=2.75]{T3}
	\Vertex[x=-1,y=2.25]{T4}
	\Vertex[x=-1,y=1.75]{T5}
	\Vertex[x=-1,y=1.25]{T6}
	\Vertex[x=-1,y=0.75]{T7}
	\Vertex[x=-1,y=0.25]{T8}
	\Vertex[x=-1,y=-0.25]{T9}
	\Vertex[x=-1,y=-0.75]{T10}
	\Vertex[x=-1,y=-1.25]{T11}
	\Vertex[x=-1,y=-1.75]{T12}
	\Vertex[x=-1,y=-2.25]{T13}
	\Vertex[x=-1,y=-2.75]{T14}
	\Vertex[x=-1,y=-3.25]{T15}
	\Vertex[x=-1,y=-3.75]{T16}
	
	\Vertex[x=1,y=3.75]{R1}
	\Vertex[x=1,y=3.25]{R2}
	\Vertex[x=1,y=2.75]{R3}
	\Vertex[x=1,y=2.25]{R4}
	\Vertex[x=1,y=1.75]{R5}
	\Vertex[x=1,y=1.25]{R6}
	\Vertex[x=1,y=0.75]{R7}
	\Vertex[x=1,y=0.25]{R8}
	\Vertex[x=1,y=-0.25]{R9}
	\Vertex[x=1,y=-0.75]{R10}
	\Vertex[x=1,y=-1.25]{R11}
	\Vertex[x=1,y=-1.75]{R12}
	\Vertex[x=1,y=-2.25]{R13}
	\Vertex[x=1,y=-2.75]{R14}
	\Vertex[x=1,y=-3.25]{R15}
	\Vertex[x=1,y=-3.75]{R16}		

	\Vertex[x=6,y=0]{X}

	\Edges(N,M1,M11)
	\Edges(N,M1,M12)
	\Edges(N,M1,M13)
	\Edges(N,M1,M14)
	
	\Edges(N,M2,M21)
	\Edges(N,M2,M22)
	\Edges(N,M2,M23)
	\Edges(N,M2,M24)
	
	\Edges(N,M3,M31)
	\Edges(N,M3,M32)
	\Edges(N,M3,M33)
	\Edges(N,M3,M34)
	
	\Edges(N,M4,M41)
	\Edges(N,M4,M42)
	\Edges(N,M4,M43)
	\Edges(N,M4,M44)	
	
	\Edges(X,H4,H41)
	\Edges(X,H4,H42)
	\Edges(X,H4,H43)
	\Edges(X,H4,H44)
	
	\Edges(X,H1,H11)
	\Edges(X,H1,H12)
	\Edges(X,H1,H13)
	\Edges(X,H1,H14)
	
	\Edges(X,H2,H21)
	\Edges(X,H2,H22)
	\Edges(X,H2,H23)
	\Edges(X,H2,H24)
	
	\Edges(X,H3,H31)
	\Edges(X,H3,H32)
	\Edges(X,H3,H33)
	\Edges(X,H3,H34)
	
	\Edges(X,H4,H41)
	\Edges(X,H4,H42)
	\Edges(X,H4,H43)
	\Edges(X,H4,H44)

	
	\Edges(M11,T5)
	\Edges(M12,T10)
	\Edges(M13,T14)
	\Edges(M14,T1)
	
	\Edges(M21,T2)
	\Edges(M22,T11)
	\Edges(M23,T15)
	\Edges(M24,T16)

	\Edges(M31,T9)
	\Edges(M32,T3)
	\Edges(M33,T12)
	\Edges(M34,T6)
	
	\Edges(M41,T4)
	\Edges(M42,T7)
	\Edges(M43,T13)
	\Edges(M44,T8)


	\Edges(M11,T16)
	\Edges(M12,T12)
	\Edges(M13,T8)
	\Edges(M14,T14)
	
	\Edges(M21,T15)
	\Edges(M22,T3)
	\Edges(M23,T2)
	\Edges(M24,T1)
	
	\Edges(M31,T13)
	\Edges(M32,T5)
	\Edges(M33,T11)
	\Edges(M34,T10)
	
	\Edges(M41,T6)
	\Edges(M42,T4)
	\Edges(M43,T7)
	\Edges(M44,T9)

	\begin{scope}
    \tikzset{EdgeStyle/.append style = {dotted}}	
	\clip(-1,3.75) rectangle (-0.6,-3.75);
	\Edges(T1,R4)
	\Edges(T2,R5)
	\Edges(T3,R14)
	\Edges(T4,R16)
	\Edges(T5,R1)
	\Edges(T6,R12)
	\Edges(T7,R11)
	\Edges(T8,R2)
	\Edges(T9,R9)
	\Edges(T10,R7)
	\Edges(T11,R10)
	\Edges(T12,R15)
	\Edges(T13,R8)
	\Edges(T14,R6)
	\Edges(T15,R3)
	\Edges(T16,R13)
	
	\Edges(T1,R14)
	\Edges(T2,R16)
	\Edges(T3,R12)
	\Edges(T4,R2)
	\Edges(T5,R11)
	\Edges(T6,R13)
	\Edges(T7,R7)
	\Edges(T8,R1)
	\Edges(T9,R3)
	\Edges(T10,R6)
	\Edges(T11,R4)
	\Edges(T12,R8)
	\Edges(T13,R10)
	\Edges(T14,R9)
	\Edges(T15,R5)
	\Edges(T16,R15)
\end{scope}
\begin{scope}
	\tikzset{EdgeStyle/.append style = {dotted}}		
	\clip(0.6,3.75) rectangle (1,-3.75);
	\Edges(T1,R15)
	\Edges(T2,R4)
	\Edges(T3,R12)
	\Edges(T4,R7)
	\Edges(T5,R9)
	\Edges(T6,R13)
	\Edges(T7,R16)
	\Edges(T8,R11)
	\Edges(T9,R8)
	\Edges(T10,R14)
	\Edges(T11,R1)
	\Edges(T12,R5)
	\Edges(T13,R10)
	\Edges(T14,R6)
	\Edges(T15,R3)
	\Edges(T16,R2)
	
	\Edges(T1,R3)
	\Edges(T2,R12)
	\Edges(T3,R6)
	\Edges(T4,R13)
	\Edges(T5,R7)
	\Edges(T6,R14)
	\Edges(T7,R1)
	\Edges(T8,R16)
	\Edges(T9,R4)
	\Edges(T10,R10)
	\Edges(T11,R2)
	\Edges(T12,R11)
	\Edges(T13,R5)
	\Edges(T14,R9)
	\Edges(T15,R8)
	\Edges(T16,R15)	
\end{scope}	

	\Edges(R1,H34)
	\Edges(R2,H11)
	\Edges(R3,H24)
	\Edges(R4,H43)
	\Edges(R5,H23)
	\Edges(R6,H22)
	\Edges(R7,H32)
	\Edges(R8,H42)
	\Edges(R9,H13)
	\Edges(R10,H12)
	\Edges(R11,H41)
	\Edges(R12,H21)
	\Edges(R13,H44)
	\Edges(R14,H31)
	\Edges(R15,H14)
	\Edges(R16,H33)
	
	\Edges(R1,H33)
	\Edges(R2,H43)
	\Edges(R3,H32)
	\Edges(R4,H44)
	\Edges(R5,H13)
	\Edges(R6,H24)
	\Edges(R7,H11)
	\Edges(R8,H41)
	\Edges(R9,H12)
	\Edges(R10,H22)
	\Edges(R11,H31)
	\Edges(R12,H42)
	\Edges(R13,H14)
	\Edges(R14,H21)
	\Edges(R15,H34)
	\Edges(R16,H23)

	\newcommand{\expanderColor}{blue}
	\begin{scope}[rotate=69]\Loop[dist=2cm,dir=WE,style={\expanderColor,-}](N)\end{scope}	
	\begin{scope}[rotate=23]\Loop[dist=2cm,dir=WE,style={\expanderColor,-}](N)\end{scope}		
	\begin{scope}[rotate=-23]\Loop[dist=2cm,dir=WE,style={\expanderColor,-}](N)\end{scope}	
	\begin{scope}[rotate=-69]\Loop[dist=2cm,dir=WE,style={\expanderColor,-}](N)\end{scope}	
	
	
	\tikzset{EdgeStyle/.style = {-,bend left=30,\expanderColor}}
	\Edges(M1,M2)
	\Edges(M2,M3)
	\Edges(M3,M4)
	\Edges(M4,M1)		
	
	\Edges(M1,M4)
	\Edges(M3,M2)
	\Edges(M3,M1)
	\Edges(M4,M2)	
	
	\tikzset{EdgeStyle/.style = {-,bend right=30,\expanderColor}}
	\Edges(T1,T8)
	\Edges(T2,T13)
	\Edges(T3,T5)
	\Edges(T4,T7)		
	\Edges(T5,T2)
	\Edges(T6,T11)
	\Edges(T7,T3)
	\Edges(T8,T9)
	\Edges(T9,T10)
	\Edges(T10,T15)
	\Edges(T11,T12)
	\Edges(T12,T16)		
	\Edges(T13,T6)
	\Edges(T14,T4)
	\Edges(T15,T14)
	\Edges(T16,T1)	
	
	\tikzset{EdgeStyle/.style = {-,bend right=30,\expanderColor}}
	\Edges(T1,T5)
	\Edges(T2,T3)
	\Edges(T3,T1)
	\Edges(T4,T15)		
	\Edges(T5,T4)
	\Edges(T6,T9)
	\Edges(T7,T14)
	\Edges(T8,T12)
	\Edges(T9,T7)
	\Edges(T10,T16)
	\Edges(T11,T13)
	\Edges(T12,T2)		
	\Edges(T13,T8)
	\Edges(T14,T10)
	\Edges(T15,T6)
	\Edges(T16,T11)		
	
	\newcommand{\decorationTreeTri}[1]
	{
		\tikzset{EdgeStyle/.style = {-,red}}	
		\Vertex[x=0.8,y=0]{Q}		
		\Vertex[x=1.6,y= 0.6]{Q1}
		\Vertex[x=1.6,y= 0.0]{Q2}	
		\Vertex[x=1.6,y=-0.6]{Q3}			
		\Vertex[x=2.4,y= 0.8]{Q11}
		\Vertex[x=2.4,y= 0.6]{Q12}
		\Vertex[x=2.4,y= 0.4]{Q13}		
		\Vertex[x=2.4,y= 0.2]{Q21}
		\Vertex[x=2.4,y= 0.0]{Q22}
		\Vertex[x=2.4,y=-0.2]{Q23}	
		\Vertex[x=2.4,y=-0.4]{Q31}
		\Vertex[x=2.4,y=-0.6]{Q32}
		\Vertex[x=2.4,y=-0.8]{Q33}			
		
		\Edges(#1,Q)
		\Edges(Q,Q1,Q11)
		\Edges(Q,Q1,Q12)		
		\Edges(Q,Q1,Q13)
		\Edges(Q,Q2,Q21)
		\Edges(Q,Q2,Q22)		
		\Edges(Q,Q2,Q23)
		\Edges(Q,Q3,Q31)
		\Edges(Q,Q3,Q32)		
		\Edges(Q,Q3,Q33)
	}	
	\newcommand{\decorationTree}[1]
	{
		\tikzset{EdgeStyle/.style = {-,red}}	
		\Vertex[x=0.8,y=0]{Q}		
		\Vertex[x=2,y= 0.8]{Q1}
		\Vertex[x=2,y= 0.6]{Q2}
		\Vertex[x=2,y= 0.4]{Q3}		
		\Vertex[x=2,y= 0.2]{Q4}
		\Vertex[x=2,y= 0.0]{Q5}
		\Vertex[x=2,y=-0.2]{Q6}	
		\Vertex[x=2,y=-0.4]{Q7}
		\Vertex[x=2,y=-0.6]{Q8}
		\Vertex[x=2,y=-0.8]{Q9}			
		
		\Edges(#1,Q)
		\Edges(Q,Q1)
		\Edges(Q,Q2)		
		\Edges(Q,Q3)
		\Edges(Q,Q4)
		\Edges(Q,Q5)		
		\Edges(Q,Q6)
		\Edges(Q,Q7)
		\Edges(Q,Q8)		
		\Edges(Q,Q9)
	}	
	\begin{scope}[shift={(X)},rotate=45]\decorationTree{X}\end{scope}	
	\begin{scope}[shift={(X)},rotate=-45]\decorationTree{X}\end{scope}	
	\begin{scope}[shift={(H4)},rotate=38]\decorationTree{H4}\end{scope}
	\begin{scope}[shift={(H4)},rotate=-10]\decorationTree{H4}\end{scope}
	\begin{scope}[shift={(T1)},rotate=135]\decorationTree{T1}\end{scope}
	\begin{scope}[shift={(T1)},rotate=45]\decorationTree{T1}\end{scope}		
					
	\node[text width=3cm, align=center] at (-6,4) {\textsf{ENTRANCE}};
	\node[text width=3cm, align=center] at (6,4) {\textsf{EXIT}};
	
	\draw [->, line width=1pt] (-6,3.5) -- (-6,.5);
	\draw [->, line width=1pt] (6,3.5) -- (6,.5);	

	\node[text width=1.5cm, align=center] at (-6,-4.25) {$C_1$};
	\node[text width=1.5cm, align=center] at (-4.5,-4.25) {$C_2$};
	\node[text width=1.5cm, align=center] at (-3,-4.25) {$C_3$};
	\node[text width=1.5cm, align=center] at (-1,-4.25) {$C_4$};
	\node[text width=1.5cm, align=center] at (0,-4.25) {$\cdots$};
	\node[text width=1.5cm, align=center] at (1,-4.25) {$C_{\ell-3}$};
	\node[text width=1.5cm, align=center] at (3,-4.25) {$C_{\ell-2}$};
	\node[text width=1.5cm, align=center] at (4.5,-4.25) {$C_{\ell-1}$};
	\node[text width=1.5cm, align=center] at (6,-4.25) {$C_\ell$};

	\draw[<->, line width=1pt] (-5.5,-5) -- (-2.5,-5);
	\node[text width=1.5cm, align=center] at (-4,-5.5) {\small Funnel};
	
	\draw[<->, line width=1pt] (-1.5,-5) -- (1.5,-5);
	\node[text width=1.5cm, align=center] at (0,-5.5) {\small Tunnel};
	
	\draw[<->, line width=1pt] (2.5,-5) -- (5.5,-5);
	\node[text width=1.5cm, align=center] at (4,-5.5) {\small Funnel};
	\end{tikzpicture}
	\caption{An illustration of a random graph that we construct for the separation. The parameters are $m=2, k=2$ and $\delta=0$ (note that these parameters are non-representative, but it is hard to draw an example with bigger parameters).
	The edges constructed during obfuscation are in black, the expander graphs on the clusters are in blue, and the decoration trees are in red. 
	For clarity of the picture we only included the expander edges on clusters $C_1,C_2$, and  $C_4$, 
	but they should be added on all clusters in our construction. Similarly, we only included decoration trees on three vertices (and limited their depth to $1$) due to space constraints, but they should be added to every vertex.}
	\label{fig:modified_glued_tree}
\end{figure}
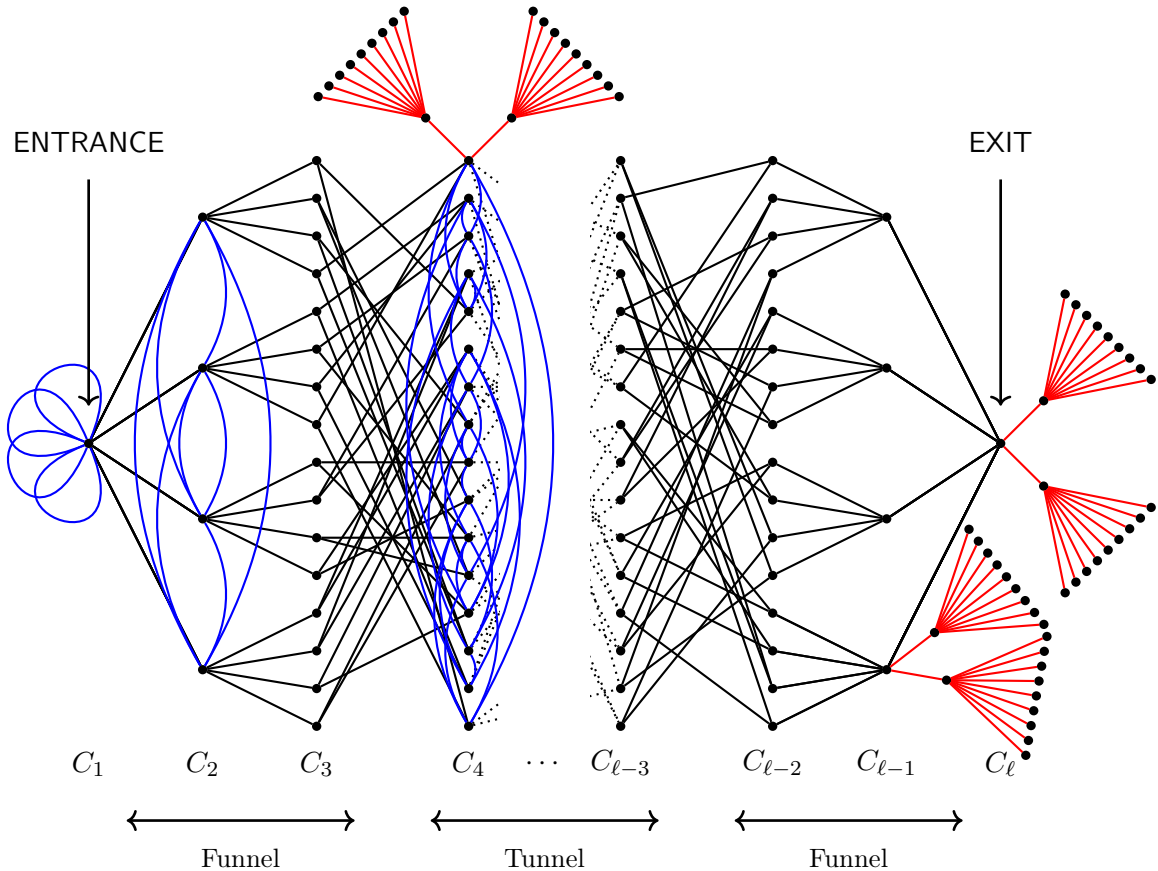

	
	The obfuscation construction is motivated by the following consideration: from the viewpoint of the quantum adiabatic algorithm, the obfuscated graph may be viewed as a path on the clusters from the ENTRANCE to the EXIT, with a weight of $m$ on each edge (the expander graph on each cluster helps enforce this structure during the adiabatic evolution). This means that for all practical purposes, the adiabatic quantum algorithm does not notice the obfuscation. On the other hand, for any classical algorithm, the obfuscated graph presents a challenge, because the graph looks locally tree-like, and the underlying path structure is effectively hidden. Nevertheless, a random walk with $\Omega(\ell^2)$ steps can still traverse the tunnel with high probability.
	
	The decoration of the obfuscated graph with trees is designed to make the graph even more difficult to navigate for any classical algorithm. Intuitively speaking the trees have a fractal-like structure with $\poly{m}$ levels of self-similarity, and each level of self-similarity will make the graph twice as difficult to navigate -- we prove this rigorously following an argument by Hastings~\cite{hastings2020PowerOfAdiabaticNoSign}. At the same time, the decoration has only an insignificant effect on the adiabatic evolution, since the spectrum of a tree of degree $d$ is bounded by $2\sqrt{d-1}$, and therefore the decoration results in only a slight $\bigO{\sqrt{m}}$-magnitude perturbation, which is too small to close the gap in the spectrum throughout the adiabatic path.  
	
	We start by stating more quantitatively the intuition that the obfuscated graph on the middle clusters is locally tree-like. After the obfuscation step the tunnel vertices all have degree~$4m$. The main observation is that for a random walk, or in fact any classical algorithm that makes only (sub)exponentially many queries, the tunnel section of the graph appears locally like a tree graph of uniform degree $4m$.
	Indeed, all matchings between the different clusters and all expander graphs within the clusters are chosen uniformly at random, so after $q$ queries the probability that a new edge query will close an inner cycle (i.e., a cycle that only contains edges within the tunnel) has probability $\bigO{q/m^k}$, so by the union bound after $q$ queries the total probability of finding a cycle is at most $\bigO{q^2/m^{k}}$. 
	
	For simplicity let us assume that $\ell$ is odd, and set $l:=(\ell-1)/2$. Consider any classical algorithm that starts at a vertex, $v$, in the middle ($(l+1)$-st) cluster; it follows that if the classical algorithm makes at most $m^{\frac{k}3}$ queries, then up to $\mathcal{O}\big(m^{-\frac{k}3}\big)$ error we can assume that the graph looks like a regular tree up to depth $l -k$. We will argue below that the decoration makes it (sub)exponentially difficult to follow a path of length $l -k$ in the original (undecorated) graph, because local exploration of the graph will be drawn into the hardly recognizable decoration trees.\footnote{We can force any classical algorithm to only do local exploration by hiding the graph among exponentially many isolated vertices, so that querying an unseen vertex label will lie outside the graph with exponentially high probability. However, this is probably not needed, as the structure itself shall conceal interesting vertices naturally, as we will see.} 
	
	It is helpful to understand the case of a single level decoration with depth-$d$ trees. Intuitively, if starting from vertex $v$, the middle section of the obfuscated graph (the tunnel) were actually a tree (instead of just looking tree-like), and if the classical algorithm was guaranteed to never make it down to a leaf of any decoration tree (with depth $d$), then we could argue as follows: from the viewpoint of the algorithm it is exploring a regular $(4m + m^{1-\delta})$-ary tree, and we are asking what is the chance that it finds a vertex that is at distance $l -k$ from $v$. In order to find such a distant vertex the algorithm has to explore at least one path of vertices of length $l -k$. The requirement of not encountering a leaf of a decoration tree forces the algorithm to stay within a (randomly embedded) subtree of degree $4m$ up to depth $l -k -d$ (that is the difference between how deep the exploration has to go and the depth of the decoration trees). Due to the random labeling of the tree this clearly fails with (sub)exponentially high probability at least $1-(1-\Theta(m^{-\delta}))^{l-k-d}$. 
	Now, of course, the tunnel is not actually a tree. But notice that the above intuition can still be made to work as follows: perform a breadth-first search from the start vertex $v$, and every time a vertex is encountered, make a new copy of it -- so that the number of vertices at depth $h$ is exactly $4m^h$. Now we can argue that from the viewpoint of the classical algorithm, the vertices at depth $l -k$ are symmetric under permutation, except when the algorithm discovers a cycle.\footnote{The same intuition is expressed slightly more rigorously in~\cite[Section IV]{childs2003ExpSpeedupQW}, where the authors say that a $q$-query classical algorithm on a regular graph can be modelled by a random embedding of a tree of size $q$.} (But as we argued above the chance of that is (sub)exponentially small.) Noting that attaching an isomorphic collection of graphs to every vertex does not change the above argument, so we get the following statement:
	\begin{lemma}[Hardness of avoiding the exploration of decoration trees]\label{lem:derailing}
		Suppose that $G$ is a rooted graph with its root $r$ having degree $k$, and all vertices up to distance $D$ have degree $k+1$. Let $G'$ be the graph where each vertex $v$ of $G$ gets $h$ distinct complete $(k+h)$-ary trees of depth $d$ attached via and edge to their root. Suppose we have access to a uniformly randomly labeled version of $G'^{(j)}$, and we can perform local exploration starting from the root $r$. Then the probability that we don't find a cycle, neither discover a leaf of any attached tree in $G'$, but discover a vertex $v$ in $G'$ at distance $D$ form $r$ has probability at most $(k/(k+h))^{D-d}$ irrespective of the number of exploration steps.
	\end{lemma}
	
	Since upon traversing the tunnel we need to go through at least one middle vertex, the above argument assures that we need to discover at least one leaf of some level-1 decoration tree, in order to traverse the tunnel. Further levels of the decoration ensure that discovering such a leaf is (sub)exponentially unlikely unless the classical algorithm makes at least (sub)exponentially many queries, as shown by the following inductive lemma. Regarding the inductive structure, note that the definition of decoration makes it possible to view $G$ as an induced subgraph of the level-$j$ decorated graph $G_j$, and in turn as an induced subgraph of the recursively decorated graph $G^{(j)}$, in particular we have $G_j^{(j-1)}= G^{(j)}$.  A we already indicated we will choose $r=m^\delta$ rounds of decoration (for simplicity let us assume that both $m^\delta$ and $m^{1-\delta}$ are integers), so that $G^{(r)}$ will look roughly uniform with degree $5m$ at every vertex around the original vertices of~$G$.\footnote{We could in principle modify the definition of decoration by adding more trees to non-maximal degree vertices, so that the resulting graph will be uniform everywhere, except at the leafs of the decoration trees. This would probably make the graph even harder to navigate for a classical algorithm. We will nevertheless stick with the above definition because it has some aspects that are more convenient for our analysis.\label{foot:improveDec}}  
	
	\begin{lemma}[Cf.~{\cite[Lemma 6]{hastings2020PowerOfAdiabaticNoSign}}]\label{lem:decHard}
	    Suppose we start form the root of a complete $(5m-jm^{(1-\delta)}-1)$-ary tree $T$ of depth $d:=(j+1) m^{(3\delta + o(1))}$ (think of $T$ as a tree attached during a $(j+1)$-level decoration), and we are only allowed to explore its decorated version $T^{(j)}$ ``locally'', i.e., by only querying neighbors of known vertices. If $j\leq m^\delta$, then for any algorithm the probability of reaching a leaf vertex of $T$ using $2^j$ queries is at most $3^{-(m^\delta -j+1)m^\delta}$.  
	\end{lemma}
	\begin{proof}
		We can prove this by induction on $j$. For $j=1$ the statement is trivial. The induction step is as follows: suppose that the statement is true for $j-1$. What we prove is that it requires at least $2^j$ queries to find a vertex that has distance at least $d$ from the root of $T$ in the graph $T^{(j)}$ with very high probability. For this we would need to traverse at least $t:=m^{3\delta + o(1)}$ edges of $T$ ($t$ is the increment in the depths of the decoration trees of subsequent levels). There are two cases:
	\begin{enumerate}[label=Case \arabic*:\kern-5mm, ref=Case \arabic*]
		\item\label{it:atm1} $\phantom{\kern5mm}$the explored vertices include leaves of at most one decoration tree in $T_j$, or
		\item\label{it:atl2} $\phantom{\kern5mm}$the explored vertices include leaves of at least two decoration trees in $T_j$.
	\end{enumerate}
	    
	First we bound the probability of \ref{it:atm1} happening. By assumption, there is a path in $T^{(j)}$ of length $d:=(j+1) m^{(3\delta + o(1))}$ going from root to leaf of $T$: all vertices on the path are explored. Furthermore, there is at most one level-$j$ decoration tree which has an explored leaf. We bound the probability of \ref{it:atm1} by case separation:
	    \begin{itemize}
	    	\item If the there is no leaf of $T_j\setminus T$ that is explored or the first explored leaf of $T_j\setminus T$ is at a distance at least $d-t/2$ from the root, then we can apply \cref{lem:derailing} to show that this event has probability at most $(1-\Theta(m^{-\delta}))^{\frac{t}{2}}$. 
	    	\item On the other hand, if there is a single decoration tree with an explored vertex whose root is at some depth at most $t/2$, then there is a path of length at least $d-t/2-1$ starting from a vertex of $v\in T$ such that all vertices along the path are explored, but no other leaf of $T_j\setminus T$ is found. We can once again apply \cref{lem:derailing} to bound the probability of this happening by $(1-\Theta(m^{-\delta}))^{\frac{t}{2}-1}$. 
	    \end{itemize}
    
	    By the union bound we get that the probability of \ref{it:atm1} is at most $2(1-\Theta(m^{-\delta}))^{\frac{t}{2}-1}\leq \exp(-m^{-2\delta+o(1)})\leq 3^{-(m^{2\delta})}$ irrespective of how many queries are made.
	    
		Now we bound the probability of \ref{it:atl2} happening. If we make at most $q$ queries, then we find at most $q$ root vertices of level-$(j-1)$ trees. For each such tree traversing to the bottom of the decoration tree takes at least $2^{(j-1)}$ queries by induction, with probability at least $1-3^{-(m^\delta -j+2))m^\delta}$. So by the union bound in order to traverse to the bottom of $2$ such trees one needs at least $2^j$ queries with probability at least $1-2^j 3^{-(m^\delta -j+2)m^\delta}$. 
		
		By applying the union bound on the distinct events \ref{it:atm1} and \ref{it:atl2} we can conclude that by using $2^j\leq 2^{m^\delta}$ queries we reach the bottom of the tree $T$ with probability at most $q 3^{-(m^\delta -j+2))m^\delta}+3^{-(m^{2\delta})}<3^{-(m^\delta -j+1)m^\delta}$.
	\end{proof}

	Consider the following three events:
	\begin{enumerate}[label=Event \arabic*:\kern-7mm, ref=Event \arabic*]
		\item\label{it:evLeaf} $\phantom{\kern7mm}$The algorithm finds a leaf of a top-level decoration tree.
		\item\label{it:evCycle} $\phantom{\kern7mm}$The algorithm finds a cycle within the tunnel.
		\item\label{it:evTraverse} $\phantom{\kern7mm}$Neither of the above two events holds but the algorithm traverses the tunnel. 		
	\end{enumerate}	

	If the algorithm finds the EXIT vertex by local exploration of the graph starting from the ENTRANCE, then it must traverse the tunnel in particular. Therefore the event of discovering the EXIT is covered by the union of the above three events. Now we bound the probability of each of the above three events, assuming that $G$ is the graph that we get by obfuscating a path of length $\ell:=m^{4\delta + o(1)}$, with $k:=3 m^{\delta}$ funnel depth, and the (classical) algorithm makes at most $q=2^{m^\delta}$ queries.
	
	Since the algorithm explores at most $q$ root vertices of a top-level decoration tree, the probability of finding a leaf of any such decoration tree with $q$ queries is at most $q3^{-m^\delta}$, due to \cref{lem:decHard} and the union bound. Therefore, the probability of \ref{it:evLeaf} is bounded by $\exp(-\Omega(m^\delta))$. 
	
	We already discussed that for any algorithm that makes at most $q$ queries the total probability of finding a cycle within the tunnel is at most $\bigO{q^2/m^{k}}$. Therefore, the probability of \ref{it:evCycle} is bounded by $\bigO{\exp(-m^\delta)}$. 
	
	Finally, if the algorithm traverses the tunnel it must reach a vertex $v$ in the tunnel from which it discovers a path of length at least $l-k$. As \cref{lem:derailing} shows, following a path of length $l-k$ from a middle vertex without discovering a cycle or a leaf of a top-level decoration is (sub)exponentially unlikely: its probability is bounded by $(1-\Theta(m^{-\delta}))^{l-k-d}$, where $d=m^{(4\delta + o(1))}$ -- the depth of the top level decoration trees. Therefore, with the right choice of the $o(1)$ term in the definition of $\ell$, we can bound the probability of \ref{it:evTraverse} by $\exp(-\Omega(m^{3\delta}))$.
	
	We can conclude using the union bound, that any classical algorithm that uses $2^{m^\delta}$ queries can reach the EXIT vertex in the decorated graph with probability at most $\exp(-\Omega(m^\delta))$. In the next subsection we will see that for preserving the gap we shall choose $\ell=\Theta( m^{\frac14})$, so we will ultimately choose $\delta=\frac{1}{16}-o(1)$. Note that the graph $G^{(0)}$ has at most $\ell m^{2k}=\bigO{\exp(m^{\delta+o(1)})}$ vertices and similarly $|G^{(m^\delta)}|\leq |G^{(0)}|\left((5m)^{m^{4\delta+o(1)}}\right)^{m^\delta}=\exp(m^{5\delta+o(1)})$. Setting $n:=m^{5\delta+o(1)}$, we can see that the vertex labels have $n$ bits, and any classical algorithm needs at least $\exp(n^{\frac{1}{5}-o(1)})$ queries to find the EXIT with probability greater than $\exp(-n^{\frac{1}{5}-o(1)})$, providing the (sub)exponential classical lower bound we claimed.

	\subsection{Preserving the adiabatic path and its main spectral properties}
	
	The actual adiabatic path that we use will be analogous to the simple path that we used at the beginning: $H(s):=(1+s)H-s H'_i$ for $s\in[-1,0]$ and $H(s):=(1-s)H+s H'_f$ for $s\in[0,1]$, where $H:=-A$, $H'_i=-m\cdot\ketbra{\mathrm{ENTRNACE}}{\mathrm{ENTRNACE}}$ and $H'_f=-m\cdot\ketbra{\mathrm{EXIT}}{\mathrm{EXIT}}$.
	
	For the sake of analysis we divide the adjacency matrix $A=A_P+A_E+A_D$ to three parts, the edges corresponding to the original path graph ($A_P$), the edges that belong to the expander graphs on the clusters ($A_E$), and the edges coming from the decoration ($A_D$). 
	
	The main idea is that for understanding the adjacency matrix after obfuscation $\tilde{A}:=A_P+A_E$ we can focus on the ``symmetric'' subspace, which is spanned by uniform superpositions $\ket{C_j}:=\frac{1}{\sqrt{|C_j|}}\sum_{v\in C_j}\ket{v}$ over the clusters $C_j\colon j\in [\ell]$. In this subspace the adjacency matrix looks like that of the original path graph $A_\ell$ just with uniform edge weights $m$. Expander graphs of uniform degree $4 m$ are added on each cluster $C_j$ corresponding to individual vertices $j\in[\ell]$ of the original path graph in order to make this ``symmetric'' subspace have the lowest energy. This enables perfectly preserving the main properties of the adiabatic path, including a spectral gap of size $\Omega(m/\ell^2)$.
	
	Indeed, let us first focus on the adiabatic path corresponding to $\tilde{A}(s):=-H(s)+(1-|s|)A_D$. $A_E$ contains no edges between clusters, but the edges form an expander with uniform degree $2 m$ within each cluster $C_j$, so that we can block-diagonalize $A_E$ according to the clusters. Clearly, then the largest eigenvalue of $A_E$ is $2 m$, and has multiplicity $\ell$, while the uniform superpositions $\ket{C_j}\colon j\in [\ell]$ form an orthonormal basis of the subspace $U$ corresponding to the largest eigenvalue, i.e.,\footnote{Here $\delta_{ij}$ stands for the Kronecker-delta.}
	\begin{equation}\label{eq:expanderGround}
		\braketbra{C_i}{A_E}{C_j}=4 m\cdot\delta_{ij}.
	\end{equation}
	With the right choice of expander graphs, the spectral gap becomes large, so that for large enough $m$ the second largest eigenvalue of $A_E$ is at most $m$, as follows from \cref{thm:expander}. Now observe that the $\ell$-dimensional subspace $U$ is also invariant under the linear map $A_P$, and that the matrix of $A_P$ on this subspace is isomorphic\footnote{Note that due to the Perron-Frobenius theorem this representation also shows that $\nrm{A_P}= m\nrm{A_\ell}\leq 2m$.}  to the adjacency matrix $A_\ell$ of the path of length $\ell$ with uniform edge-weights $m$, i.e.,\footnote{This can be seen as follows: if $S$ and $T$ are disjoint subsets of the vertices of a graph $G$ with $M$ edges between $S$ and $T$, then for $\ket{S}:=\frac{1}{\sqrt{|S|}}\sum_{v\in S}\ket{v}$ and $\ket{T}:=\frac{1}{\sqrt{|T|}}\sum_{w\in T}\ket{w}$ we have that  $\braketbra{S}{A_G}{T}=\frac{M}{\sqrt{|S|\cdot|T|}}$.\label{foot:symmMatElements}}
	\begin{equation}\label{eq:pathIsom}
		\braketbra{C_i}{A_P}{C_j}=m \cdot \braketbra{i}{A_\ell}{j}.
	\end{equation}
	Since $U$ is also invariant under $\tilde{A}(s)$, we can see that our analysis of the spectrum of the simple adiabatic path $H'(s)$ directly applies here as well, and so we get that the spectral gap within $U$ has size $\Omega(m/\ell^2)$.
	
	Now we show that the second largest eigenvalue of $\tilde{A}(s)$ comes from the spectrum on $U$.
	Let $\bar{U}$ denote the orthogonal complement of $U$, and $\Pi_{\bar{U}}$ denote the orthogonal projection to it. 
	The largest absolute eigenvalue of $\tilde{A}$ on the invariant subspace $\bar{U}$ can be written as $\nrm{\Pi_{\bar{U}}\tilde{A}\Pi_{\bar{U}}}$, which can be bounded by $\nrm{\Pi_{\bar{U}}A_P\Pi_{\bar{U}}}+\nrm{\Pi_{\bar{U}}A_E\Pi_{\bar{U}}}\leq \nrm{A_P}+m$, therefore the largest eigenvalue of $\tilde{A}(s)$ on $\bar{U}$ is at most $(1-|s|)(\nrm{A_P}+m)\leq (1-|s|)3m$. At the same time, our analysis in \cref{sec:pathGap} shows that the second largest eigenvalue of $A_\ell(\alpha)\geq 1$ for $\ell\geq 5$ (cf. \cref{fig:evSolutionsPlot}) and thereby the second largest eigenvalue of $\tilde{A}(s)$ on $U$ is at least $(1-|s|)3m$. Thus, we can conclude that the overall spectral gap of $\tilde{A}(s)$ has size $\Omega(m/\ell^2)$.
	
	The second main idea is that the graph of edges added during decoration form a forest of maximum degree $5m$ and thus the corresponding adjacency matrix $A_D$ has spectral norm bounded\footnote{This is straightforward to show, for example using the argument of Hastings~\cite{hastings2020PowerOfAdiabaticNoSign}: a forest of degree at most $d+1$ can be embedded into a uniform tree that has $d$ children at every level for some finite depth $t$. Due to the Perron-Frobenius theorem the largest eigenvalue of the forest can be bounded by the largest eigenvalue of the tree, which can again be bounded by $2\sqrt{d}$. The largest eigenvalue of the uniform tree is easy to bound by reducing it to the path graph of length $t$ with edge weights $\sqrt{d}$; again it suffices to considering the ``symmetric'' subspace similarly to our previous argument.} by $2\sqrt{5m}$. Therefore, as long as $\ell\ll m^{\frac14}$, perturbation by the matrix $(|s|-1)A_D$ cannot close\footnote{This is again straightforward to show using an argument by Hastings~\cite{hastings2020PowerOfAdiabaticNoSign}: if the Hermitian matrix $B$ has a spectral gap $3\gamma$ around its largest eingenvalue, and the Hermitian matrix $C$ has norm at most $\gamma$, then $B+C$ has a spectral gap at least $\gamma$. Indeed, let $\lambda$ be the largest eigenvalue and $\ket{\psi}$ the corresponding eigenvector of $B$, then the largest eigenvalue of $B+C$ can be lower bounded by $\braketbra{\psi}{B+C}{\psi}\geq \lambda - \gamma$. At the same time by the Courant-Fischer-Weyl min-max principle we have that the second largest eigenvalue of $B+C$ is at most $\max_{\phi}\braketbra{\phi}{(I-\ketbra{\psi}{\psi})(B+C)(I-\ketbra{\psi}{\psi})}{\phi}\leq
	\max_{\phi}\braketbra{\phi}{(I-\ketbra{\psi}{\psi})B(I-\ketbra{\psi}{\psi})}{\phi}
	+\max_{\phi}\braketbra{\phi}{(I-\ketbra{\psi}{\psi})C(I-\ketbra{\psi}{\psi})}{\phi}\leq
	\lambda-3\gamma +\gamma=\lambda-2\gamma$.} the gap of $\tilde{A}(s)$ which has size $\Omega(m/\ell^2)$. Therefore, choosing $\ell=\Theta(m^{\frac14})$ appropriately ensures that the adiabatic path $H(s)$ has a spectral gap of size at least $\Omega(m/\ell^2)$ around its ground state energy. 
	
	The above two main observations show that the adiabatic path still maps the ENTRANCE vertex to the EXIT vertex. Since the main Hamiltonian comes from the adjacency matrix of an undirected and unweighted graph	 it has no sign problem.
	
	\section{Relation to the welded-trees construction and quantum walks}\label{sec:weldedWalk}
	Our graph is reminiscent to the welded-trees construction~\cite{childs2003ExpSpeedupQW}. The main difference from the welded-trees construction is that our graph has highly non-uniform degree distribution. This is necessary for getting a polynomially large overlap between the ENTRANCE and EXIT vertices, and the largest eigenvalue of the adjacency matrix. (Indeed, a uniform degree-$d$ graph has largest eigenvalue $d$, and the corresponding eigenvector is a uniform superposition over all vertices.) Therefore the analysis of~\cite{childs2003ExpSpeedupQW} that heavily relied on the uniformity of the welded-trees graph, does not apply here, and a different construction was necessary. Our modification is very natural: simply weld the trees with less edges, in order to get the sought polynomial overlap with the ENTRANCE and EXIT vertices. But there is a difficulty here arising from the fact that the non-uniform degrees can give away the structure of the trees, allowing fast traversal~\cite{childs2020GraphPropExpQSpeedup,benDavid2020SymmetriesGraphPropertiesQSpeedups}. A first attempt is to add a longer middle section, or tunnel, between the trees --- but unfortunately a random walk can still traverse such a tunnel. Nevertheless, the tunnel is uniform and looks locally tree-like, and we can utilize these properties to hide the edges of the original graph, by adding ``camouflaged'' decoration trees, motivated by the recent work of Hastings~\cite{hastings2020PowerOfAdiabaticNoSign}. It is worth noting that our new construction is much more similar to the welded-trees construction~\cite{childs2003ExpSpeedupQW} than that of Hastings~\cite{hastings2020PowerOfAdiabaticNoSign}.
	
	The similarity to the welded-trees construction hints at the possibility to use a natural quantum walk algorithm in addition to the adiabatic path that we described. Since we have adjacency-list access to the graph, we can implement an efficient \emph{block-encoding}~\cite{gilyen2018QSingValTransf} of its adjacency matrix divided by the maximal degree $(m^2+m+1)$ (i.e., a quantum circuit which corresponds to a unitary matrix $U$ whose top-left corner equals $A/(m^2+m+1)$), using just $2$ queries. Such a block-encoding can be used for running a Szegedy-type discrete quantum walk~\cite{szegedy2004QMarkovChainSearch,childs2008OnRelContDiscQuantWalk,ambainis2019QuadSpeedupFindingMarkedQW}. Let $\ket{\psi}$ denote the top eigenvector of $A$; we can approximately ``project'' this block-encoding to a block-encoding $U'$ of $\ketbra{\psi}{\psi}$ via quantum singular value transformation~\cite{low2017HamSimUnifAmp,gilyen2018QSingValTransf} according to a low-degree $\bigOt{m^{\frac{3}{2}}}$ polynomial approximation of the threshold function filtering out all non-maximal eigenvalues~\cite{gilyen2018QSingValTransfArXiv,lin2019OptimalQEigenstateFiltering}; this $U'$ can be implemented using $\bigOt{m^{\frac{3}{2}}}$ queries. Applying this block-encoding to $\ket{\mathrm{ENTRNACE}}$ results in a polynomially large overlap with $\ket{\mathrm{EXIT}}$, since $\bra{\mathrm{EXIT}}\ketbra{\psi}{\psi}\ket{\mathrm{ENTRNACE}}=\left|\braket{\psi}{\mathrm{ENTRNACE}}\right|^2=\Omega(\ell^{-3})=\Omega(m^{-\frac{3}{4}})$. Using amplitude amplification this gives an $\bigOt{m^{\frac{9}{4}}}$-time quantum query algorithm.
	
	Alternatively, the above approximate threshold polynomial can be decomposed into a linear combination of Chebyshev polynomials, where the coefficients have $\ell^1$-weight at most $\bigOt{m^{\frac{3}{4}}}$. Since the Szegedy quantum walk effectively applies a Chebyshev polynomial to the block-encoded adjacency matrix~\cite{childs2008OnRelContDiscQuantWalk,gilyen2018QSingValTransfArXiv,ambainis2019QuadSpeedupFindingMarkedQW}, it enables a linear combination of unitaries (LCU)~\cite{childs2012HamSimLCU,childs2015QLinSysExpPrec,gilyen2018QSingValTransf} based implementation of an $\widetilde{\Omega}(m^{\frac{3}{4}})$-subnormalized block-encoding of $\ketbra{\psi}{\psi}$. This algorithm can be simplified~\cite{apers2019UnifiedFrameworkQWSearch} by randomly picking a time $t\in[\widetilde{\mathcal{O}}(m^{3/2})]$, and applying $t$-steps of the Szegedy walk to the input state $\ket{\mathrm{ENTRNACE}}$. It is not difficult to show~\cite{apers2019UnifiedFrameworkQWSearch} that measuring the final state of this plain quantum-walk-based algorithm finds the EXIT with probability $\Omega(1/\poly{m})$. 	
	
	\section{Discussion}
	In this paper we have demonstrated the possibility of a (sub)exponential quantum speedup via an adiabatic evolution where the instantaneous Hamiltonians have spectral norm at most $\poly{n}$ and the spectral gap is at least $1/\poly{n}$, furthermore the Hamiltonian has no sign problem, i.e., all of its matrix elements are non-positive. In order to prove such a big separation we worked in an oracle model, where a $\poly{n}$-sparse graph is given via its adjacency list, whose adjacency matrix described the Hamiltonian. The adiabatic path has very nice additional properties: the initial and the final Hamiltonians are diagonal, and the path consists of two ``straight lines''; in fact only following the first ``line segment'' already provides the (sub)exponential quantum speedup, but then the final Hamiltonian is non-diagonal. Our result heavily builds on ideas recently introduced by Hastings~\cite{hastings2020PowerOfAdiabaticNoSign}, but simplifies and improves his result in several ways: Hastings' result only showed a superpolynomial quantum speed-up, and used significantly more complicated Hamiltonian paths. Additionally, our problem features a natural quantum walk algorithm, providing a new example of a (sub)exponential speedup via a simple quantum walk.
	
	In order get a (sub)exponential quantum speedup, we used a graph where the top eigenvector's $\ell^2$-weight is concentrated on the essential structural parts of the graph, whereas the $\ell^1$-weight is concentrated on some ``camouflaging decorations''. This $\ell^1$-weight shift already happened after one level of decoration which intuitively ruled out efficient simplistic Monte Carlo algorithms. In order to rule out any efficient classical algorithm we added a polynomial number of decoration layers, effectively hiding the essential structure of the graph from any classical algorithm.
	
	We conjecture that it should be possible to improve the exponent of our (sub)exponential lower bound to $\exp(n^{1-o(1)})$ for some simple adiabatic path with no sign problem. In fact, we think that a fine-tuned version of the construction discussed in this paper might already exhibit such a separation. We already hinted at the possibility of some improved decoration structure in \cref{foot:improveDec}, and there are other potential improvements that shall improve the exponent. For example the use of the union bound over all encountered roots inside the proof of \cref{lem:decHard} is probably unnecessary, and ultimately one might not need to increase the depth of the decoration trees between various levels as rapidly as we do in \cref{def:dec}. 
	
	The big open question that remains is whether one can get a superpolynomial speedup via an adiabatic path that has no sign problem, and comes from a local Hamiltonian. Such a speedup could have important practical implications, since D-Wave's quantum annealers have such restrictions. However, proving such a result requires proving a superpolynomial circuit lower-bound for a non-oracular problem, which is beyond the reach of currently known techinques in theoretical computer science.
	
	\section*{Acknowledgments}
	Part of this work was done while visiting the Simons Institute for the Theory of Computing; we gratefully acknowledge its hospitality.
	
	A.G.\ acknowledges funding provided by Samsung Electronics Co., Ltd., for the project ``The Computational Power of Sampling on Quantum Computers'', and additional support by the Institute for Quantum Information and Matter, an NSF Physics Frontiers Center (NSF Grant PHY-1733907). U.V.\ was supported by the Vannevar Bush faculty fellowship N00014-17-1-3025, and NSF QLCI Grant No. 2016245.
	
	\bibliographystyle{alphaUrlePrint}
	\bibliography{Bibliography}
	
	\appendix
	
	\section{The spectral gap of the adiabatic evolution on the path graph}\label{sec:pathGap}
	
	In this appendix we prove that the energy gap around the ground state of $H_\ell(s)$ is at least $\Omega(1/\ell^2)$ for all $s\in[-1,1]$. Since $\nrm{H_\ell(s)}=\Theta(1)$, this is equivalent to proving that $H_\ell(s)/\nrm{H_\ell(s)}$ has an $\Omega(1/\ell^2)$-large gap around its ground state energy. We prove the latter, but using a more convenient parametrization.
	
	Consider the Hamiltonian $A_\ell(\alpha):= \alpha \ketbra{0}{0} + A_\ell$, for $\alpha\in \mathbb{R}_+$. We will show that the smallest eigenvalue gap of $A_\ell(\alpha)$ is at least $\Omega(1/\ell^2)$. This will imply that the spectral gap of $A_\ell(\alpha)$ around the largest eigenvalue is $\Omega((1+\alpha)/\ell^2)$, which in turn implies that the spectral gap around the ground state energy of $-A_\ell(\alpha)/\nrm{A_\ell(\alpha)}$ is at least $\Omega(1/\ell^2)$ proving that $H'(s)$ has an $\Omega(1/\ell^2)$-large gap around its ground state energy.
	
	Now we turn to analyzing $A_\ell(\alpha)$. We claim that the eigenvectors and eigenvalues of $A_\ell(\alpha)$ are all associated to solutions of the quasimomenta equation
	\begin{equation}\label{eq:quasiMomenta}
	f_\ell(p):=\frac{\sin((\ell+1)p)}{\sin(\ell p)}=\alpha.
	\end{equation}
	Indeed the vector $\ket{\psi_p}:=\sum_{j=1}^{\ell}\sin(j p)\ket{j}$ is always an eigenvector of $A_\ell(\alpha)$ with eigenvalue $2\cos(p)$, whenever $p$ is a solution of \cref{eq:quasiMomenta}. Since $f_\ell(p)$ and $\cos(p)$ are both symmetric and $2\pi$ periodic it suffices to concentrate on solutions within the interval $[0,\pi]$.
	
	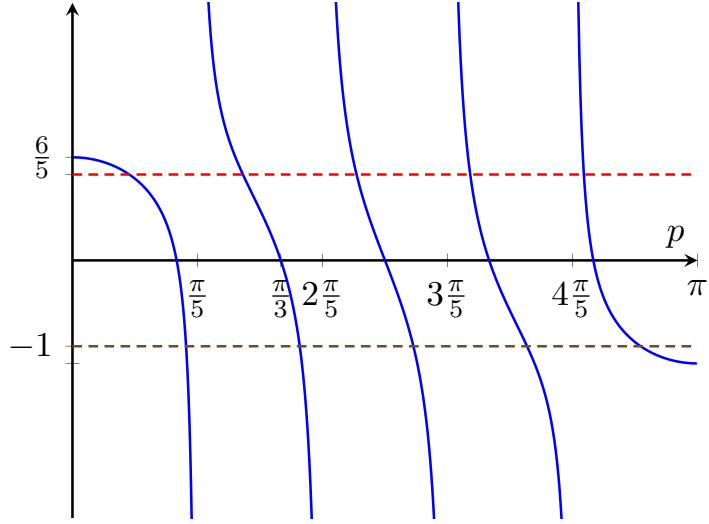
\begin{figure}[H]
		\begin{center}
			\begin{tikzpicture}[scale=1.2]
			\begin{axis}[xlabel={$p$}, axis lines=middle, samples=120, domain=0:
			pi, ymin=-3, ymax=3,
			thick, trig format plots=rad,no markers,
			xtick = {0,0.2*pi,0.333*pi,0.4*pi,0.6*pi,0.8*pi,pi},
			xticklabels = {$0$,$\frac{\pi}{5}$,$\frac{\pi}{3}$,$2\frac{\pi}{5}$,$3\frac{\pi}{5}$,$4\frac{\pi}{5}$,$\pi$},
			ytick = {-1.2,-1,0,1,1.2},
			yticklabels = {$ $,$-1$,$0$,$ $,$\frac{6}{5}$},
			]
			\addplot[blue,domain=0.001*pi:0.199*pi] {sin(6*x)/sin(5*x)};
			\addplot[blue,domain=0.201*pi:0.399*pi] {sin(6*x)/sin(5*x)};		
			\addplot[blue,domain=0.401*pi:0.599*pi] {sin(6*x)/sin(5*x)};	
			\addplot[blue,domain=0.601*pi:0.799*pi] {sin(6*x)/sin(5*x)};	
			\addplot[blue,domain=0.801*pi:0.999*pi] {sin(6*x)/sin(5*x)};				
			\addplot {1};		
			\addplot {-1};		
			\end{axis}
			\end{tikzpicture}
		\end{center}
		\caption{
			Plot of $f_\ell(p)\!=\!\frac{\sin((\ell+1)p)}{\sin(\ell p)}$ for $\ell\!=\!5$; dashed lines show the solutions of \cref{eq:quasiMomenta} for $\alpha\!=\!\pm 1$.
		}\label{fig:evSolutionsPlot}
	\end{figure}
	
	We show that for $\alpha\in [0,\frac{\ell+1}{\ell})$ there are $\ell$ distinct $p\in[0,\pi]$ solutions to $\cref{eq:quasiMomenta}$, and there are $\ell-1$ such solutions otherwise. For $\alpha=\frac{\ell+1}{\ell}$ the additional eigenvalue is $2$ which can be obtained as the limit of the largest eigenvalue as $\alpha$ goes to $\frac{\ell+1}{\ell}$ from below and corresponds to the ``pseudo-solution'' $p=0+$ corresponding to the eigenvector $\ket{\psi_{0+}}:=\sum_{j=1}^{\ell}j\ket{j}$. If $\alpha > \frac{\ell+1}{\ell}$, then there is a complex solution to \cref{eq:quasiMomenta}, which corresponds to the unique real solution $x$ of $\frac{\sinh((\ell+1)x)}{\sinh(\ell x)}=\alpha$ giving eigenvalue $2\cosh(x)$ and eigenvector $\ket{\phi_x}:=\sum_{j=1}^{\ell}\sinh(j x)\ket{j}$.
		
	We proceed by showing that on every interval of the form $(\frac{j-1}{\ell}\pi,\frac{j}{\ell}\pi)$ for $j\in [\ell]$ the function $f_\ell(p)$ is strictly monotone decreasing, and the range of $f_\ell(p)$ equals $\mathbb{R}$ on these intervals apart from the first and last intervals. For this observe that for all $j\in[\ell-1]$
	\begin{equation*}
		\lim_{\eps\downarrow 0} f_{\ell}\left(\frac{j}{\ell}\pi\pm \eps\right)
		=\lim_{\eps\downarrow 0} \left(\frac{\sin(j\pi +\frac{j}{\ell}\pi\pm(\ell+1)\eps)}{\sin(j \pi\pm \ell\eps)}\right)
		=\lim_{\eps\downarrow 0} \left(\frac{\sin(\frac{j}{\ell}\pi\pm(\ell+1)\eps)}{\sin(\pm\ell\eps)}	\right)	
		=\pm\infty.
	\end{equation*}
	Further observe that the derivative 
	\begin{equation*}
	f'_\ell(p)=f_\ell(p)((\ell+1)\cot((\ell+1)p)-\ell\cot(\ell p))
	\end{equation*}
	is non-positive for all $p\in[0,\pi]\setminus \mathbb{N}\frac{\pi}{\ell}$. To see this, observe that $f_\ell(p)$ changes sign where either $\sin(\ell p)$ or $\sin((\ell+1)p)$ changes sign, that is at values $p\in S_\ell:=\{\frac{j}{\ell}\pi\colon j\in[\ell-1]\}\cup\{\frac{j}{\ell+1}\pi\colon j\in[\ell]\}$. Now we show that $g_\ell(p):=((\ell+1)\cot((\ell+1)p)-\ell\cot(\ell p))$ also changes sign at exactly the same set of points. 
	Since $\ell \cot(\ell p)$ jumps from $-\infty$ to $\infty$ at $p\in \{\frac{j}{\ell}\pi\colon j\in[\ell-1]\}$ and $(\ell + 1)\cot((\ell + 1) p)$ jumps from $-\infty$ to $\infty$ at $p\in \{\frac{j}{\ell+1}\pi\colon j\in[\ell]\}$ we get that $g_\ell(p)$ changes sign at $p\in S_\ell$. One can also see that $g_\ell(p)\neq 0$ for any $p\in(0,\pi)$ implying that $g_\ell(p)$ changes sign only at points of $S_\ell$. Indeed
	\begin{align*}
	g_\ell(p)&=0\\
	&\Updownarrow \\
	\cot((\ell+1)p)&=\ell(\cot(\ell p)-\cot((\ell+1)p))\\
	&\Downarrow \\[-5mm]
\cos((\ell+1)p)&=\ell\frac{\overset{\sin(p)}{\overbrace{\sin((\ell+1)p)\cos(\ell p)-\cos((\ell+1)p)\sin(\ell p))}}}{\sin(\ell p)},
	\end{align*}
	where the last equality clearly does not hold, since $\cos((\ell+1)p)\in[-1,1]$, whereas in contrast $|l\sin(p)|/|\sin(\ell p)|>1$ for all $p\in(0,\pi)$. This implies that $f'_\ell(p)=f_\ell(p)g_\ell(p)$ is never positive. 
	
	We can conclude that $f_\ell(p)$ is strictly monotone decreasing on the intervals $(\frac{j-1}{\ell}\pi,\frac{j}{\ell}\pi)$ for all $j\in[\ell]$, moreover the range of $f_\ell(p)$ equals $\mathbb{R}$ on these intervals for all $j\in\{2,3,\ldots,\ell-1\}$, while the range of $f_\ell(p)$ on $(0, \frac{\pi}{\ell})$ equals $(-\infty,\frac{\ell+1}{\ell})$ and its range on $(\frac{ \ell-1}{\ell}\pi,\pi)$ equals $(\frac{\ell+1}{\ell},\infty)$.
	This proves our claim about the number of solutions of \cref{eq:quasiMomenta} within the interval $(0,\pi)$.
	
	Next, we prove our lower bound on the spectral gap. Firs we show that for any $\alpha\in\mathbb{R}$ the $\ell$ or $\ell-1$ different real solutions of \cref{eq:quasiMomenta} have an $\Omega(1/\ell)$ gap in between. This implies that the corresponding eigenvalues also have gaps of size at least $\Omega(1/\ell^2)$ due to the following little lemma:
	\begin{lemma}
		Suppose that $x<y\in[0,\pi]$, then $|\cos(x)-\cos(y)|\geq 1-\cos(y-x)\geq (y-x)^2/\pi^2$.
	\end{lemma}
	\begin{proof}
		\begin{align*}
			\!\cos(x)\!-\!\cos(y)
			=\!\int_x^y \kern-2mm\sin(z) dz 
			\geq \! \int_0^{y-x}\kern-6mm \sin(z) dz 
			= \cos(0)\!-\!\cos(y\!-\!x) 
			= 1\!-\!\cos(y\!-\!x)\geq \frac{(y\!-\!x)^2}{\pi^2}. & \qedhere
		\end{align*}
	\end{proof}

	For lower bounding the gaps between the solutions to \cref{eq:quasiMomenta}, observe that for $\alpha=1$ the solutions are $\{\frac{2j-1}{2\ell +1}\pi\colon j\in [\ell]\}$, and similarly for $\alpha=-1$ the solutions are $\{\frac{2j}{2\ell +1}\pi\colon j\in [\ell]\}$. Since $f_\ell(p)$ is strictly monotone decreasing within each interval $[\frac{(j-1)}{\ell}\pi,\frac{j}{\ell}\pi]$ for all $j\in[\ell]$ we get that for any $\alpha\in[-1,1]$ the $j$-th solution of \cref{eq:quasiMomenta} lies in the interval $[\frac{2j-1}{2\ell +1}\pi,\frac{2j}{2\ell +1}\pi]$. Thus for every $\alpha\in[-1,1]$ any two subsequent solutions have a gap at least $\frac{2j+1}{2\ell+1}\pi-\frac{2j}{2\ell+1}\pi=\frac{\pi}{2\ell+1}$. Similarly for $|\alpha|>1$ the solutions lie outside the intervals $[\frac{2j-1}{2 \ell+1}\pi,\frac{2j}{2 \ell+1}\pi]\colon j\in[\ell]$ and are therefore also at least $\frac{\pi}{2\ell+1}$ apart. We can conclude that the different real solutions of \cref{eq:quasiMomenta} have gaps of size at least $\frac{\pi}{2\ell+1}$ in between. 

	We also need to treat the case when $\alpha\geq \frac{\ell+1}{\ell}$, so that there are only $(\ell-1)$ real solutions to \cref{eq:quasiMomenta}. then the largest eigenvalue is $2\cosh(x)$ for some $x\in \mathbb{R}$. Moreover, $\braketbra{1}{A_\ell(\alpha)}{1}= \alpha$, so we get that the largest eigenvalue is at least $\max\{2,\alpha\}$. On the other hand the second largest eigenvalue is $2\cos(p)$ for some $p\in(\frac{1}{\ell}\pi,\frac{2}{\ell}\pi)$, and since $2\cos(p)< 2-p^2/\pi^2$ for all $p\in(0,\pi)$, 
	we get that the second largest eigenvalue is at most $2-1/\ell^2$. Thus the eigenvalue gap is at least $\max\{1/\ell^2,\alpha-2+1/\ell^2\}=\Omega((\alpha+1)/\ell^2)$.

	Thus we have shown that for any $\alpha\geq 0$ the matrix $A_\ell(\alpha)$ has a spectral gap at least $\Omega((\alpha+1)/\ell^2)$. Since $\nrm{A_\ell(\alpha)}\leq \alpha + \nrm{H_\ell}=\bigO{\alpha+1}$ we get that the spectral gap of $A_\ell(\alpha)/\nrm{A_\ell(\alpha)}$ is at least $\Omega(1/\ell^2)$ finishing our proof of the fact that $H_\ell(s)$ has an $\Omega(1/\ell^2)$-large gap around its ground state energy for every $s\in[-1,0]$, and due to symmetry this result extends to $s\in[-1,1]$.
	
	Finally, let us understand the solutions and eigenvalues for the unperturbed case when $\alpha=0$. Clearly the $\ell$ different solutions of \cref{eq:quasiMomenta} are $\{\frac{j}{\ell+1}\pi\colon j\in[\ell]\}$. In this case the largest eigenvalue of $A_\ell(0)=A_\ell$ is $2\cos(1/(\ell\!+\! 1))$, and the corresponding eigenstate is proportional to $\sum_{j=1}^\ell \sin(\frac{j}{\ell+1}\pi)\ket{j}$. In particular the normalized eigenstate has an overlap of at least $\Omega(\ell^{-\frac32})$ with any vertex $\ket{j}$.
	
	\section{The effect of decoration on the top eigenvector: \texorpdfstring{$\ell^1$ vs. $\ell^2$ weight}{l1 vs. l2 weight}}\label{apx:l1vsl2}
	In this appendix we study the effect of decoration on the top eigenvector of the adjacency matrix. 
	We first show that for any graph $G$ with top-eigenvector $\psi$, the top eigenvector $\psi'$ of the decorated graph $G'$ is proportional to $\psi$ on the original vertices of $G$. (This result also applies to the Hamiltonians that come from intermediate $s\neq 0$ Hamiltonians.) Moreover, we show that in the top eigenvector $\psi^{(r)}$ of $G^{(r)}$ -- the graph that we construct in this paper -- the overwhelming majority of the $\ell^2$-weight is supported on the original vertices of $G$. On the other hand the $\ell^1$-weight proportion of the original vertices of $G$ in $\psi^{(r)}$ is (sub)exponentially small.
	
	Suppose we have a graph $G=(V,E)$, and we attach $k$-copies of a connected graph $T$ to every vertex of $G$ via an edge to a distinguished vertex $t$ of $T$ resulting in the new graph $G'$. 
	Let $\lambda_G$ be the top eigenvalue of $G$ with corresponding eigenvector $\psi$. For $\gamma\in\mathbb{R}_+$ let $T(\gamma)$ be the graph where we add a self-loop to the vertex $t$ with weight $\gamma$ and let $\lambda_T(\gamma)$ its largest eigenvalue with $\phi(\gamma)$ the corresponding eigenvector normalized such that the amplitude $\phi_t(\gamma)$ at $t$ equals $1$. Then we claim that the top eigenvalue $\lambda_{G'}$ equals $\phi(\gamma)$, where $\gamma$ is the unique solution\footnote{Since the right-hand side of \cref{eq:decorEV} is strictly monotone increasing for $\gamma\in\mathbb{R}_+$ and the left-hand side is strictly monotone decreasing there is at most one solution. Since both sides are continuous, and in the $\gamma\rightarrow 0$ limit the left hand side is $+\infty$ and in the $\gamma\rightarrow +\infty$ limit the left-hand side is $+\infty$ there always must be a solution to \cref{eq:decorEV}.} to the equation 
	\begin{equation}\label{eq:decorEV}
	\lambda_G+\frac{k}{\gamma} = \lambda_T(\gamma).
	\end{equation}
	It is easy to verify that the corresponding eigenvector is $\psi'=\psi+ \bigoplus_{v\in V}\frac{\psi_v}{\gamma} \bigoplus_{j \in [k]} \phi(\gamma)$, where $\bigoplus_{j \in [k]} \phi(\gamma)$ stands for the direct sum of the eigenvectors $\phi(\gamma)$ corresponding to the $k$ copies of $T$ attached to the vertex $v$.
	
	Now that we have an analytic description of how the eigenvectors and eigenvalues change under decoration it is time to understand the quantity $\lambda_T(\gamma)$, when $T$ is a complete tree with $d=\Theta(m)$ children at every level up to depth $\ell'$ for some $\ell'\in \{m^{3\delta},m^{3\delta}+1,\ldots, m^{4\delta}\}$, and $w$ is the root of $T$ like in our decorations. In our scenario we have $2m\leq\lambda_G$, $k=m^{1-\delta}$, and for large enough $m$
	\begin{equation}\label{eq:gammaBound}
		m\leq \gamma\leq  \lambda_G+1.
	\end{equation}
	To see the latter, consider \cref{eq:decorEV} and observe that $\lambda_T(m)\leq m + 2\sqrt{d}=m+o(m)$ which is smaller than $\lambda_G$ for large enough $m$, while $\lambda_T(\lambda_G+1)\geq \lambda_G+1$ which is larger than $\lambda_G+k/\gamma\leq \lambda_G+\bigO{m^{-\delta}}$ for large enough $m$. 
	
	It is again useful to consider the adjacency matrix of $T(\gamma)$ in the ``symmetric'' subspace, which is spanned by uniform superpositions $\ket{L_j}=\frac{1}{\sqrt{|L_j|}}\sum_{v\in L_j}\ket{v}$ over the level sets of $T$ $L_j=\{\text{vertices of }T\text{ at distance }j\text{ form the root}\}$. The symmetric subspace is again invariant under the linear map $A_T(\gamma)$, and due the Perron-Frobenius theorem it contains the largest eigenvector, moreover its matrix looks exactly like the adjacency matrix of a path graph of length-$\ell'$ with a self-loop of weight $\gamma$ at the root, and uniform $\sqrt{d}$ edge weights, cf. \cref{foot:symmMatElements}. When $\gamma \geq 2\sqrt{d}$ by our analysis of such graphs in \cref{sec:pathGap} we know that the largest eigenvalue will be $\sqrt{d}\cosh(x)$, where $x$ is the unique solution of the equation 
	\begin{equation}\label{eq:decorPathEval}
		\sinh((\ell'+2)x)=\frac{\gamma}{\sqrt{d}}\sinh((\ell'+1)x),
	\end{equation}
	and the corresponding eigenvector is proportional to $\sum_{j=0}^{\ell'}\sinh((\ell'+1-j)x)\ket{L_j}$. Thus, the normalized eigenvector $\phi(\gamma)$ can be written as
	\begin{equation}\label{eq:decorPathEvec}
	\phi(\gamma)=\sum_{j=0}^{\ell'}\frac{\sinh((\ell'+1-j)x)}{\sinh((\ell'+1)x)}\ket{L_j}.
	\end{equation}
	
	Let us now bound the $\ell^2$-weight of $\phi(\gamma)$ as follows
	\begin{equation}\label{eq:l2weightbound}
		\nrm{\phi(\gamma)}_2
		\leq\sum_{j=0}^{\ell'}\left|\frac{\sinh((\ell'+1-j)x)}{\sinh((\ell'+1)x)}\right|\nrm{\ket{L_j}}_2
		= \sum_{j=0}^{\ell'}\prod_{i=0}^{j-0}\frac{\sinh((\ell'-i)x)}{\sinh((\ell'+1-i)x)}
		\leq \sum_{j=0}^{\ell'}\left(\frac{\sqrt{d}}{\gamma}\right)^{\! j},
	\end{equation}	
	where the last inequality follows from \cref{eq:decorPathEval} and the fact\footnote{This can be seen by 
		$-\cosh(2x)\leq -1\Rightarrow$
		$e^{2hx}+e^{-2hx} - e^{2x}-e^{-2x}\leq e^{2hx}+e^{-2hx} -2 \Rightarrow$
		$4 \sinh((h+1)x)\sinh((h-1)x)\leq 4\sinh^2(hx) \Rightarrow$
		$\frac{\sinh((h-1)x)}{\sinh(hx)}\leq\frac{\sinh(hx)}{\sinh((h+1)x)}$.} that $\frac{\sinh((h-1)x)}{\sinh(hx)}$ is monotone increasing in $h$ for every $x\geq0$. 
	Since $\frac{\sqrt{d}}{\gamma}=\bigO{m^{-\frac{1}{2}}}$, \cref{eq:l2weightbound} implies that $\nrm{\phi(\gamma)}_2=\Theta(1)$. Then the $\ell^2$-weight of $\nrm{\frac1\gamma\bigoplus_{j \in [k]} \phi(\gamma)}_2=\frac{\sqrt{k}}{\gamma}\nrm{\phi(\gamma)}_2=\bigO{m^{-\frac{1+\delta}{2}}}$. Therefore, after one level of decoration the $\ell^2$-weight ratio of the original eigenvector $\psi$ within the new eigenvector $\psi'$ is $1-\bigO{m^{-(1+\delta)}}$, and after $m^\delta$ levels of decoration the ratio still remains as large as $1-\bigO{1/m}$.
	
	Finally, let us bound the $\ell^1$-weight of $\phi(\gamma)$ by observing that the solution of \cref{eq:decorPathEval} satisfies $x\leq\ln(\gamma/\sqrt{d})$.\footnote{Our analysis of \cref{eq:quasiMomenta} revealed that \cref{eq:decorPathEval} has at most $1$ real solution. Then the function $\sinh((\ell'+2)x)/\sinh((\ell'+1)x)$ must be strictly monotone increasing on $\mathbb{R}_+$ due to its continuous behavior. Since for every $c>1$ we have $\sinh((\ell'+2)\ln(c))/\sinh((\ell'+1)\ln(c))=c (1-c^{-2(\ell'+2)})/(1-c^{-2(\ell'+1)})>c$ we get $x\leq \ln(\gamma/\sqrt{d})$.}
	\begin{align}
		\nrm{\phi(\gamma)}_1
		&=\sum_{j=0}^{\ell'}\frac{\sinh((\ell'+1-j)x)}{\sinh((\ell'+1)x)}\nrm{\ket{L_j}}_1\nonumber\\
		&\geq\sum_{j=0}^{\ell'}\frac{\sinh((\ell'+1-j)x)}{2\exp((\ell'+1)x)}\nrm{\ket{L_j}}_1\nonumber\\
		&=\sum_{j=0}^{\ell'}\frac{\exp((\ell'+1-j)x)}{\exp((\ell'+1)x)}(1-\exp(-2(\ell'+1-j)x))\nrm{\ket{L_j}}_1\nonumber\\
		&\geq(1-\exp(-2x))\sum_{j=0}^{\ell'}\frac{\exp((\ell'+1-j)x)}{\exp((\ell'+1)x)}\nrm{\ket{L_j}}_1\nonumber\\	
		&=(1-\Theta(d/\gamma^2))\sum_{j=0}^{\ell'}\exp(-jx)d^{\frac{j}{2}}\nonumber\\	
		&\geq(1-\Theta(d/\gamma^2))\sum_{j=0}^{\ell'}\left(\frac{d}{\gamma}\right)^{\!j}.\label{eq:l1lowerbound}
	\end{align}
	In our case the obfuscated graph satisfies $\lambda_G< 4m$, and for the first decoration we have $d=4m+m^{1-\delta}$ so that $\frac{d}{\gamma}=1+\Omega(m^{-\delta})$, therefore \cref{eq:l1lowerbound} implies $\nrm{\phi(\gamma)}_1=\exp(\Omega(m^{2\delta}))$. Thus, already after one level of decoration the $\ell^1$-weight ratio of the original eigenvector $\psi$ within the new eigenvector $\psi'$ is $\exp(-\Omega(m^{2\delta}))$. Since later decorations also satisfy $\frac{d}{\gamma}=1+\Omega(m^{-\delta})$, after applying all $m^\delta$ levels of decoration the ratio becomes as small as $\exp(-\Omega(m^{3\delta}))$.
	
	Finally, note that essentially the same argument as above shows that the $\ell^2$-weight of the top eigenvector is concentrated on the original vertices of the obfuscated graph throughout the entire adiabatic path. However, the above argument breaks down for showing that only a tiny a fraction of the $\ell^1$-weight is located on the original graph. Indeed, for $|s|=1$ the top eigenstate is supported on the original graph, and due to continuity it implies that for $|s|\approx 1$ most of the $\ell^1$-weight is located on the original graph.

	\section{Sampling random regular expander graphs}\label{sec:expanders}

	\begin{definition}[Sampling a random cycle]\label{def:cycleSamp}
		For any  $n\in \mathbb{N}$ we can sample a uniformly random cycle on the vertex set $[n]$ as follows: sample an arbitrary permutation of $n$ elements, and define the corresponding edge set as $\{(i,j)\in [n]\times [n]\colon \pi(i)-\pi(j) \equiv \pm 1 \mod n \}$.
	\end{definition}
		Note that there is an alternative way to sample uniformly random cycles which is equivalent to the above:
		sample a cyclic permutation $\pi$ of $[n]$ independently and uniformly at random, and define the set of edges so that $(i,j)\in E$ iff $\pi^{\pm 1}(i)=j$. However, in our analysis \cref{def:cycleSamp} is more convenient.
	
	\begin{definition}[Random $d$-regular graphs]\label{def:randReg}
		For an even $d$ an $n\in \mathbb{N}$ we denote by $\mathcal{H}_{n,d}$ a distribution of undirected $d$-regular graphs $G=(V,E)$ on $n$ vertices. Identifying the vertices with the set $V:=[n]$ the distribution is defined by independently sampling $d/2$ random cycles as in \cref{def:cycleSamp}, and taking the union of their edges (with multiplicity).
	\end{definition}
	
	\begin{theorem}[{\cite[Theorem 1.2]{friedman2003ProofAlonSecondEVConj}}]\label{thm:expander}
		Fix a real $\eps > 0$ and an even positive integer $d$. Then there is a constant, $c$, such that for a random graph, $G$, in $\mathcal{H}_{n,d}$ we have that with probability at least $1 - c/n^\tau$ the second largest eigenvalue  $\lambda_2(A_G) \leq 2 \sqrt{d - 1} + \eps$, where $\tau = \tau_{\mathrm{fund}} = \lceil \sqrt{d - 1} \rceil - 1$.
	\end{theorem}
	
	\begin{corollary}\label{cor:ourExpanders}
		There is a universal constant $C$, such that for every $m,n\in\mathbb{N}$ satisfying $n\geq m^2$ we have that a random $G$ sampled from $\mathcal{H}_{n,2m}$ satisfies $\lambda_2(A_G) \leq m$ with probability at least $1-\frac{C}{m^5}$.
	\end{corollary}
	\begin{proof}
		Let $k:=\lfloor m/8\rfloor$ and $r:=m-8k$ for some $k\in\mathbb{N}$. Then we can obtain a sample $G$ by taking $k$ independent samples $H_i$ from $\mathcal{H}_{n,16}$, and a sample from $H_0$ from $\mathcal{H}_{n,2r}$ and taking the union of their edges (with multiplicity). Setting $\eps=0.1$ and $d=16$ in \cref{thm:expander}, and observing that $2\sqrt{15}+0.1< 7.9$, we get that there is some universal constant $c\in\mathbb{R}_+$ such that for each $i\in[k]$
		\begin{equation}\label{eq:expand16}
		\Pr(\lambda_2(H_i)> 7.9)\leq cn^{-3}\leq cm^{-6}.
		\end{equation}
		By the union bound we get that with probability at least $1- c m^{-5}$ \cref{eq:expand16} holds for all $i\in[k]$ simultaneously, and therefore we get that $\Pr(\lambda_2(G)> 7.9 k + 2r=m + r-0.1k)\leq cm^{-5}.$ If $m\geq 560$, then $r\leq 7\leq 0.1k$, so for such large $m$ we also get $\Pr(\lambda_2(G)> 2m)\leq cm^{-5}.$ Choosing $C:=\max(560^5,c)$ establishes the statement of the theorem.
 	\end{proof}

\end{document}